\documentclass[reqno,12pt,letterpaper]{amsart}
\usepackage{amsmath,amssymb,amsthm,graphicx,mathrsfs,url}
\usepackage[usenames,dvipsnames]{color}
\usepackage[colorlinks=true,linkcolor=Red,citecolor=Green]{hyperref}
\usepackage{amsxtra}
\usepackage{placeins}
\usepackage{wasysym} % for "\varhexagon" macro
\usepackage{graphicx}% for "\scalebox" macro
\usepackage{subcaption}
\usepackage{array}
\newcolumntype{P}[1]{>{\centering\arraybackslash}m{#1}}

\def\?[#1]{\textbf{[#1]}\marginpar{\Large{\textbf{??}}}}

\def\smallsection#1{\smallskip\noindent\textbf{#1}.}
\let\epsilon=\varepsilon 

\newtheorem{prop}{Proposition}[section]	

\newtheorem{conj}{Conjecture}

\newtheorem{lemm}[prop]{Lemma}

\numberwithin{equation}{section}

\newtheorem*{theo*}{Theorem}

\DeclareMathOperator{\Spec}{Spec}

\title[Semiclassical Band Splitting]
{Semiclassical band splitting in the infinite Chern band model}

\author{Tyler Guo}
\email{tylerguo@berkeley.edu}
\address{Department of Mathematics, University of California, Berkeley, CA 94720}
\begin{document}
\begin{abstract}
   We study the two lowest Bloch bands of the infinite Chern band model of Tan--Devakul \cite{tri} in the semiclassical limit. When the Berry curvature parameter is an odd multiple of \(\pi\),
   we show that the first two bands are exponentially close and meet at
   exactly one Dirac point. The proof combines the approach of \cite{GuZw26} with Agmon estimates and WKB approximation for an effective matrix valued Hamiltonian. The results and their potential generalizations are illustrated numerically.
      %we derive an explicit asymptotic formula for their exponentially small splitting, uniformly over the Brillouin torus. The proof combines Agmon estimates and WKB approximations for a matrix Schrödinger operator whose lowest potential eigenvalue has two wells. Geometric properties of the Agmon distance allow us to compute the leading tunneling coefficients by the method of stationary phase. We provide numerical computations to illustrate the asymptotic formula and support the conjecture that the two bands remain separated for other values of the Berry curvature parameter.
\end{abstract}
\maketitle

% \begingroup
% \setlength{\parskip}{0pt}
% \linespread{1}\small
% \tableofcontents
% \par\removelastskip
% \vspace{0pt}
% \endgroup

\section{Introduction}
We study the lowest bands of the \emph{infinite Chern band model}
introduced by Tan--Devakul \cite[Section~II.C.2]{tri}.
% \medskip\noindent
% \begin{minipage}{\textwidth}
% \centering
\begin{figure}
\includegraphics[width=15cm]{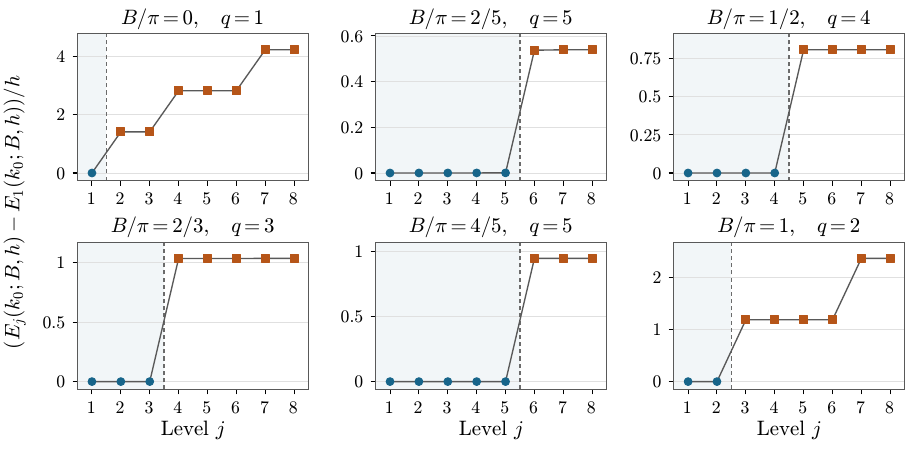}
\captionsetup{hypcap=false}
\captionof{figure}{Rescaled energy jumps for $ h = 0.01$
showing the exponential proximity depending on $ B = 2 \pi p/q$. We conjecture
that the bands are nevertheless separated for $ B \notin ( 2 \ell + 1 )
\pi$.}
%levels
%$(E_j(M;B,h)-E_1(M;B,h))/h$ at $M=(1/2,1/2)$ and $h=0.01$.
% The Fourier indices satisfy $-63\leq m,n\leq64$.
% Blue circles mark $j\leq q$ and orange squares mark $j>q$; the dashed
% line separates the groups. Vertical scales differ.
% Values visually equal to zero need not vanish.}
\label{app:fig-rational}
\end{figure}
%\end{minipage}
Motivated by quantum anomalous Hall effects in rhombohedral multilayer graphene, Tan–Devakul \cite{tri} introduced this idealized model to study how parent Berry curvature influences the topology of states induced by strong repulsive interactions or a periodic potential. We consider the noninteracting setting, in which a periodic potential splits the parent band into Bloch bands that can carry nonzero Chern numbers. Numerical computations in \cite[Section~III, Figure~2(b)]{tri} show successive closings of the gap between the first two bands as $B$ varies, accompanied by changes in the Chern number of the lowest band. These observations
motivate a precise analysis of the separation and touching
of the first two bands.

 In the differential-operator formulation, the model is given by a Hamiltonian acting on
$L^2(\mathbb{R}^2_x\times\mathbb{R}_w)$:
\begin{equation}\label{fullhamiltonian}
\begin{gathered}
\mathcal{H}_J
=\mathcal{D}^*\mathcal{D}
+J(A-\lambda\mathcal{D})^*(A-\lambda\mathcal{D})
+V_0(\cos x_1+\cos x_2),\\
\mathcal{D}=D_{x_1}+iD_{x_2},
\qquad
A=\frac{1}{\sqrt{2}}(D_w-iw).
\end{gathered}
\end{equation}
Here $D=\frac{1}{i}\partial$, $V_0$ is the amplitude of the periodic
potential, and $J$ sets the excitation energy of the internal
oscillator. The parameter $B=2\lambda^2$ is the magnitude of the uniform
Berry curvature of the parent band, defined as the lowest band of the Hamiltonian with $V_0=0$.

A mathematical treatment of the Hamiltonian \eqref{fullhamiltonian} was provided in \cite{GuZw26}.
In particular, \cite[Theorem~1]{GuZw26} proves that,
as $J\to\infty$, the Bloch eigenvalues of
\eqref{fullhamiltonian} below a fixed energy are approximated,
with error $\mathcal{O}(J^{-1})$, by those of
\begin{equation}\label{effectivehamiltonian}
\mathcal{H}_{\gamma,B}
=-\Delta+\gamma\bigl(\cos(x_1+BD_{x_2})+\cos x_2\bigr),
\qquad
\gamma=V_0e^{-B/4}.
\end{equation}
The factor $e^{-B/4}$ comes from averaging the potential against the oscillator ground state. %We study this effective operator after the large-$J$ limit.

For small $\gamma>0$, the lowest bands for general $B$ are analyzed
using perturbation theory in \cite[Theorem~2]{GuZw26}. In particular,
the minimum gap $g_{\gamma}(B)$ between the first two bands of
$\mathcal{H}_{\gamma,B}$ satisfies
\begin{equation}\label{smallcouplinggap}
g_\gamma(B)
=\gamma\bigl|\,|\cos(B/4)|-|\sin(B/4)|\,\bigr|
+\mathcal{O}(\gamma^2).
\end{equation}
The leading term in \eqref{smallcouplinggap} vanishes at odd
multiples of $\pi$. At these values
of $B$, symmetry forces exact degeneracy at
$k_{0}=(\frac{1}{2},\frac{1}{2})$ for all $\gamma>0$, and the crossing
is conical for sufficiently small $\gamma>0$
\cite[Theorem~3]{GuZw26}. Although conical crossings persist
generically under continuation in $\gamma$, this does not ensure
that the crossings remain in the lowest pair of bands.

In this paper, we address the complementary regime $\gamma\to\infty$,
setting $h=\gamma^{-1/2}$ and rescaling the fiber operator
associated with \eqref{effectivehamiltonian} by $\gamma^{-1}$:
\begin{equation}\label{originaloperator}
\begin{split}
H(k):=(hD_{x_{1}}-hk_{1})^{2}+(hD_{x_{2}}&-hk_{2})^{2}+\cos(x_{1}+B(D_{x_{2}}-k_{2}))+\cos x_{2}
\\
B\in (2\mathbb{Z}+1)\pi,\ \  \ &(x_{1}, x_{2})\in\mathbb{R}^{2}/2\pi\mathbb{Z}^{2}, \ \ (k_{1}, k_{2})\in\mathbb{R}^{2}/\mathbb{Z}^{2}.
\end{split}
\end{equation}
Our main result shows that the first two bands are exponentially
close and meet at exactly one Dirac point.
\begin{theo*}\label{maintheorem}
Let $E_{1}(k)\leq E_{2}(k)\leq\cdots$ be the eigenvalues of $H(k)$, counted with multiplicity. There exist constants $ C, S>0$, independent of $h$ and $k$, such that, as $h\rightarrow 0$
\begin{equation}\label{globalasymptotics}
 \begin{split}
     E_{2}(k)-E_{1}(k)&=Ch^{\frac{1}{2}}e^{-S/h}\sqrt{\cos^{2}(\pi k_{1})+\cos^{2}(\pi k_{2})}+\mathcal{O}(h^{\frac{3}{2}}e^{-S/h})
 \end{split}   
\end{equation}
uniformly for $k\in\mathbb{R}^{2}/\mathbb{Z}^{2}.$ Moreover, for every sufficiently small fixed $h>0$, the point $k_{0}=(\frac{1}{2}, \frac{1}{2})$ is a Dirac point of the first two bands. Writing $E_{D}(h):=E_{1}(k_{0})=E_{2}(k_{0}),$ we have
\[
\begin{gathered}
E_{1,2}(k_0+\kappa)
=E_D(h)\mp\alpha(h)|\kappa|+\mathcal O_h(|\kappa|^2),
\qquad \kappa\longrightarrow0,\\
\alpha(h)=\tfrac{1}{2}C\pi\,h^{1/2}e^{-S/h}
\bigl(1+\mathcal O(h)\bigr)>0.
\end{gathered}
\]
\end{theo*}
The tunneling action $S$ is given explicitly as
\[
S:=\int_{0}^{\pi}\sqrt{\sqrt{2}-\sqrt{1+\cos^{2}t}}\,\mathrm{d}t.
\]
The accuracy of the approximation is illustrated by
Figure~\ref{app:fig-pi} below. We refer to the \hyperref[Appendix]{appendix} for the numerical value of $C$.

\begin{figure}[!ht]
\centering
\includegraphics[width=\textwidth]{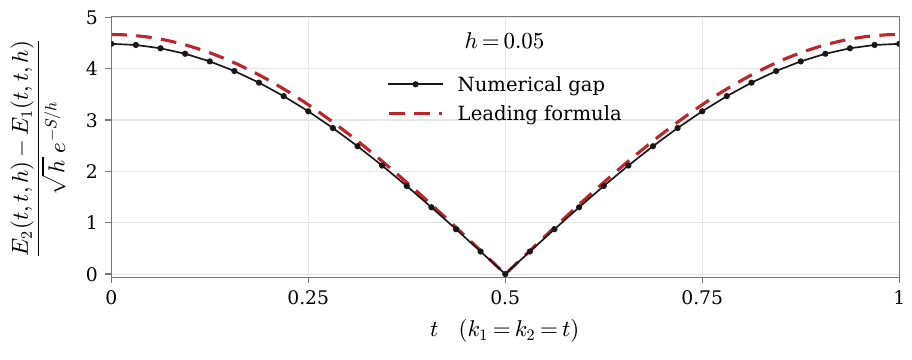}
\caption{Diagonal section at $h=0.05$ and $B=\pi$: numerical
$(E_2(t,t)-E_1(t,t))/(\sqrt h\,e^{-S/h})$ (black) and leading
profile $C\sqrt2|\cos(\pi t)|$ (red dashed).
Click \href{https://band-geometry-supplement.fiberinfinity.chatgpt.site/}
{here} for an interactive three-dimensional plot.}
\label{app:fig-pi}
\end{figure}
%\FloatBarrier

Our proof of the main theorem is based on the semiclassical framework for multi-well problems developed by Helffer--Sj\"ostrand \cite{HeSj84,HeSj85}, and its periodic implementation by Outassourt \cite{Ou87}. The novelty lies in applying this framework to a specific model of current interest in condensed matter physics, taking advantage of its special structure. We place the model in the broader setting of Schrödinger operators on Hermitian vector bundles, and give a coordinate-free formulation of Agmon estimates and the WKB construction. In the study of Agmon estimates we emphasize
the origin of the weight as a viscosity solution of the eikonal equation. A further contribution of the paper is a detailed analysis of the matrix potential arising from this model. We establish several geometric properties of the Agmon distance that enable explicit calculations of the leading profile of the splitting.

In Section~\ref{review}, we study low-energy eigenfunctions of semiclassical Schrödinger operators on Hermitian vector bundles over a closed Riemannian manifold $(M, g)$. We prove Agmon estimates that provide exponential decay of low-energy eigenfunctions away from the zero set of the lowest eigenvalue $\mu$ of the matrix potential. The decay rate is quantified by the distance function associated with the Agmon metric $\mu g.$ When the lowest eigenvalue $\mu$ is simple near a nondegenerate minimum, we construct semiclassical quasimodes with phase given by the Agmon distance from that minimum and leading amplitude taking values in the corresponding eigenspace.

In Section~\ref{3}, we reduce \eqref{originaloperator} to a $2\times 2$ matrix Schr\"odinger operator whose lowest potential eigenvalue has two nondegenerate minima. Comparison with one-well Dirichlet realizations
separates the first two bands from the remaining spectrum. Using localized one-well eigenfunctions on the universal cover, we construct, for each $k$, an orthonormal basis for the corresponding spectral subspace and derive an asymptotic formula for the effective matrix. Its eigenvalues determine the band splitting in terms of exponentially small coefficients. The leading
coefficients are expressed through surface integrals on small hypersurfaces transverse to the minimizing Agmon geodesics joining neighboring lifted wells. We then apply WKB approximations and the method of stationary phase near the intersection points, where the Agmon distances are smooth and their sum has a nondegenerate minimum on each hypersurface. The non-vanishing of the leading amplitudes then gives nonzero leading coefficients.

A natural further problem is to describe the lowest bands for
general $B$ in the semiclassical limit. If $B/\pi\in\mathbb{Q}$,
a similar reduction to a differential operator with a matrix-valued
potential is available; Appendix~\ref{app:rational-B} outlines how
this yields exponential clustering of the lowest bands.
For irrational $B/\pi$, the same
finite-rank reduction does not apply. It would be interesting to develop a semiclassical method valid
without a rationality assumption and, in particular, to determine
whether the first two bands remain separated for each fixed
$B\notin(2\mathbb{Z}+1)\pi$ as $h\to0$.

\smallsection{Acknowledgments} We thank Maciej Zworski for several helpful discussions. We gratefully acknowledge partial support from the Simons Foundation through Targeted Grant Award No. 896630, “Moir\'e Materials Magic”.

\smallsection{AI disclosure}
The main body of the paper was developed and written by the author. ChatGPT 6 was used to draft the numerical appendix and to generate figures following detailed instructions. AI was also used to identify symmetries specific to the model that simplify the proof of Proposition \ref{prop:dirac}, as well as for editorial suggestions and computational verifications. The author reviewed and revised the generated material and takes full responsibility for the paper.
\section{Semiclassical preliminaries}\label{review} 
\par We let $(M, g)$ be a closed, connected and oriented manifold, $E\rightarrow M$ be a finite-rank Hermitian vector bundle. Let $\nabla:C^{\infty}(M;E)\rightarrow C^{\infty}(M;T^{*}M\otimes E)$ be a metric connection, and $V\in C^{\infty}(M;\mathrm{End}(E))$ be a smooth Hermitian endomorphism.
\par
We adapt the presentations from \cite[Chapters 3 and 6]{DiSj99} to the semiclassical Schr\"odinger operator $P_{h}\in\mathrm{Diff}^{2}(M;E)$ of the form
\begin{equation}\label{eq2.1}
P_{h}:=h^{2}\nabla^{*}\nabla+V.    
\end{equation}
We assume \begin{equation}\label{eigenvalueassumption}
 \mu(x):=\min(\Spec V|_{E_{x}})\geq 0,\ \    \Gamma:=\{x\in M\mid \mu(x)=0\}\neq\emptyset,
\end{equation}
and regard $P_{h}$ as the self-adjoint operator on $L^{2}(M;E)$ with domain $H^{2}(M;E).$
\subsection{Agmon estimates}
We show exponential decay of low-energy eigenfunctions of $P_{h}$ within $\{x\in M\mid \mu(x)>0\}.$

We let $\widetilde{g}:=\mu g$ be the Agmon metric and define the Agmon distance function by
\begin{equation}\label{Agmondistancefunction}
\begin{split}
\ell_{\widetilde{g}}(\gamma)&:=\int_{0}^{1}\sqrt{\mu(\gamma(t))}|\dot{\gamma}(t)|_{g}\,\mathrm{d}t,
\\
 \widetilde{\Phi}(x)&:=\inf_{\gamma}\Big\{\ell_{\widetilde{g}}(\gamma)\Big|\, \gamma(0)\in\Gamma,\ \gamma(1)=x, \gamma\in C^{\infty}([0, 1];M)\Big\}.  
 \end{split}
\end{equation}
\begin{prop}\label{exponentialdecay}
    Suppose $P_{h}$ is given by \eqref{eq2.1} and satisfies the assumptions \eqref{eigenvalueassumption}. If $u(h)$ satisfies
 \begin{equation}\label{normalizedeigenfunction}
P_{h}u(h)=\lambda(h)u(h),\ \ \|u(h)\|_{L^{2}(M;E)}=1,\ \ \lambda(h)\xrightarrow{h\rightarrow 0} 0,     
 \end{equation}
 then for every $\epsilon>0$, there exist $C_{\epsilon}>0$ and $h_{\epsilon}>0$ such that for $0<h<h_{\epsilon},$
 \begin{equation}\label{agmonestimateI}
 \|e^{\widetilde{\Phi}/h}u(h)\|_{L^{2}(M;E)}+\|\nabla(e^{\widetilde{\Phi}/h}u(h))\|_{L^{2}(M;T^{*}M\otimes E)}\leq C_{\epsilon}e^{\epsilon/h}.    
 \end{equation}
\end{prop}
We first provide  generalizations of \cite[Propositions 6.1 and 6.2]{DiSj99} to our setting.
\begin{lemm}\label{agmon1}
    Let $u\in C^{\infty}(M; E)$, $\Phi\in W^{1,\infty}(M;\mathbb{R})$, and $P_{h}$ be given in \eqref{eq2.1}. Then we have
    \begin{equation}\label{agmonidentity}
    \begin{split}
        &    \mathrm{Re}\int_{M}e^{2\Phi/h}\big\langle P_{h}u,u\big\rangle_{E_{x}}\,dV_{g}\\
        & \ \ \ \ \ \ \ \ \ =h^{2}\|\nabla (e^{\Phi/h}u)\|^{2}_{L^{2}(M; T^{*}M\otimes E)}+\int_{M}e^{2\Phi/h}\big\langle(V-|d\Phi|^{2}_{g})u, u\big\rangle_{E_{x}}\,dV_{g}.    \end{split}
    \end{equation}
\end{lemm}
\begin{proof}
    We start with $\Phi\in C^{\infty}(M;\mathbb{R})$ and define $A, B:C^{\infty}(M;E)\rightarrow C^{\infty}(M;T^{*}M\otimes E)$ by
    \[
    A:=h\nabla, \ \ B:=d\Phi\otimes\bullet.
    \]
    A simple computation gives
    \[
 e^{\Phi/h}Ae^{-\Phi/h}=A-B,\ \ \ e^{\Phi/h}A^{*}e^{-\Phi/h}=(A+B)^{*}.   
    \]
    If we take $v:=e^{\Phi/h}u\in C^{\infty}(M; E)$, then \eqref{agmonidentity} follows from the identity
    \[
    \begin{split}
    h^{2}\mathrm{Re}\big\langle e^{\Phi/h}\nabla^{*}\nabla e^{-\Phi/h}v, v\big\rangle_{L^{2}(M; E)}&=\mathrm{Re}\big\langle (A-B)v, (A+B)v\big\rangle_{L^{2}(M; T^{*}M\otimes E )}
    \\
    &=\|Av\|_{L^{2}(M;T^{*}M\otimes E)}^{2}-\|Bv\|_{L^{2}(M; T^{*}M\otimes E)}^{2}
    \\
    &=h^{2}\|\nabla v\|_{L^{2}(M; T^{*}M\otimes E)}^{2}-\int_{M}|d\Phi|_{g}^{2}\,|v|^{2}_{E_{x}}\,dV_{g}.
\end{split}
    \]
    Now suppose $\Phi\in W^{1,\infty}(M;\mathbb{R}).$ Choose a finite atlas $(U_{j},\kappa_{j})$ and a partition of unity $(\chi_{j})$ with $\mathrm{supp} \chi_{j}\Subset U_{j}.$ Convolving each $(\chi_{j}\Phi)\circ\kappa_{j}^{-1}$ using a standard mollifier $\eta_{\epsilon}$ gives a sequence (see \cite[Theorem 4.1]{EvGa15}) $\{\Phi_{\epsilon}\}\subseteq C^{\infty}(M)$ with
    \[
    \|\Phi_{\epsilon}-\Phi\|_{L^{\infty}(M)}\rightarrow 0, \ \ \sup_{\epsilon}\|d\Phi_{\epsilon}\|_{L^{\infty}(M)}<\infty, \ \ d\Phi_{\epsilon}\rightarrow d\Phi\ \mathrm{a.e.}
    \]
    Hence, \eqref{agmonidentity} extends to $\Phi\in W^{1,\infty}(M;\mathbb{R})$ by the dominated convergence theorem.
    \end{proof}
    \begin{lemm}\label{agmonii}
        Under the assumptions of Lemma \ref{agmon1}, let $F_{+}, F_{-}\in L^{\infty}(M;[0,\infty))$ satisfy
        \begin{equation}\label{agmoniieq}
         \big\langle (V-|d\Phi|^{2}_{g})w, w\big\rangle_{E_{x}}\geq \big(F_{+}^{2}(x)-F^{2}_{-}(x)\big)|w|^{2}_{E_{x}} \ \ \mathrm{a.e.}
        \end{equation}
        Then
        \begin{equation}\label{agmoniieq2}
        \begin{split}
        h^{2}\|\nabla(e^{\Phi/h}u)\|^{2}_{L^{2}(M;T^{*}M\otimes E)}+\tfrac{1}{2}\|F_{+}e^{\Phi/h}u\|^{2}_{L^{2}(M; E)}
        \ \ \ \ \ \ \ \ \\\leq\|\tfrac{1}{F_{+}+F_{-}}e^{\Phi/h}P_{h}u\|^{2}_{L^{2}(M;E)}+\tfrac{3}{2}\|F_{-}e^{\Phi/h}u\|^{2}_{L^{2}(M; E)}. 
        \end{split}
        \end{equation}
    \end{lemm}
    \begin{proof}
        From \eqref{agmonidentity} and \eqref{agmoniieq} we obtain
        \begin{equation*}
            \begin{split}
               & h^{2}\|\nabla(e^{\Phi/h}u)\|^{2}_{L^{2}(M;T^{*}M\otimes E)}+\|F_{+}e^{\Phi/h}u\|^{2}_{L^{2}(M;E)} \\
               & \ \ \ \ \leq \|\tfrac{1}{F_{+}+F_{-}}e^{\Phi/h}P_{h}u\|_{L^{2}(M;E)}\|(F_{+}+F_{-})e^{\Phi/h}u\|_{L^{2}(M;E)} +\|F_{-}ue^{\Phi/h}\|^{2}_{L^{2}(M; E)}
                \\
                & \ \ \ \ \leq \|\tfrac{1}{F_{+}+F_{-}}e^{\Phi/h}P_{h}u\|^{2}_{L^{2}(M;E)}+\tfrac{1}{2}\|e^{\Phi/h}F_{+}u\|^{2}_{L^{2}(M; E)}
               +\tfrac{3}{2}\|e^{\Phi/h}F_{-}u\|^{2}_{L^{2}(M; E)}.
            \end{split}
        \end{equation*}
\end{proof}
\noindent\emph{Proof of Proposition \ref{exponentialdecay}.}
 1. We show that $\widetilde{\Phi}$ defined by \eqref{Agmondistancefunction} satisfies
\begin{equation}\label{viscosity}
  |d\widetilde{\Phi}|^{2}_{g}=\mu\ \ \mathrm{a.e.}  
\end{equation}
From \cite[Theorem 4.5 and 6.5]{EvGa15} and the inequality
\[
|\widetilde{\Phi}(x)-\widetilde{\Phi}(y)|\leq d_{\widetilde{g}}(x, y)\leq \|\sqrt{\mu}\|_{L^{\infty}(M)}\,d_{g}(x, y),
\]
we conclude that $\widetilde{\Phi}$ is differentiable almost everywhere. Suppose now $x\notin \Gamma$ and $\widetilde{\Phi}$ is differentiable at $x.$ Then
\[
|d\widetilde{\Phi}(x)|_{g}=\limsup_{y\rightarrow x}\frac{|\widetilde{\Phi}(y)-\widetilde{\Phi}(x)|}{d_{g}(x, y)}\leq\lim_{r\rightarrow 0}\sup_{z\in B_{g}(x, r)}\sqrt{\mu(z)}=\sqrt{\mu(x)}.
\]
For the reverse inequality, let $r>0$ be sufficiently small so that $\overline{B_{g}(x, r)}\cap \Gamma=\emptyset$. Choose a curve $\gamma_{r}$ from $x$ to $\Gamma$ satisfying
\[
\ell_{\widetilde{g}}(\gamma_{r})\leq \widetilde{\Phi}(x)+r^{2}.
\]
Let $y_{r}$ be the first point at which $\gamma_{r}$ meets $\partial B_{g}(x, r).$ Then
\[
|d\widetilde{\Phi}(x)|_{g}\geq \limsup_{r\rightarrow 0}\frac{\widetilde{\Phi}(x)-\widetilde{\Phi}(y_{r})}{d_{g}(x,y_{r})}\geq\limsup_{r\rightarrow 0}\big(\inf_{z\in \overline{B_{g}(x,r)}}\sqrt{\mu(z)}-r\big)=\sqrt{\mu(x)}.
\]
If $x\in \Gamma$ and $\widetilde{\Phi}$ is differentiable at $x$, then $|d\widetilde{\Phi}(x)|=\sqrt{\mu(x)}=0.$

 \noindent 2. In order to use Lemma \ref{agmonii} to control the weighted eigenfunction away from $\Gamma$, we need $F_{+}$ given in \eqref{agmoniieq} to have a positive lower bound away from a neighborhood of $\Gamma$. We therefore need to take $0<\alpha<1$ and $\Phi_{\alpha}:=(1-\alpha)\widetilde{\Phi}.$  We have for a.e. $x\in M,$
\[
\begin{split}
\big\langle (V-\lambda(h)-|d\Phi_{\alpha}|^{2}_{g})w, w\big\rangle_{E_{x}}&\geq \big((2\alpha-\alpha^{2})\mu(x)-\lambda(h)\big)|w|^{2}_{E_{x}}
\\
&=\big(F^{2}_{+,\alpha,h}(x)-F^{2}_{-,\alpha,h}(x)\big)|w|^{2}_{E_{x}}
\end{split}
\]
with
\begin{equation}\label{F_pm}
F_{\pm,\alpha, h}(x):=\sqrt{\big((2\alpha-\alpha^{2})\mu(x)-\lambda(h)\big)_{\pm}}.    
\end{equation}
Let $v_{\alpha,h}:=e^{\Phi_{\alpha}/h}u.$ Equation \eqref{agmonidentity} shows that
\begin{equation}\label{proofofproposition2.1}
 h^{2}\|\nabla v_{\alpha,h}\|^{2}_{L^{2}(M; T^{*}M\otimes E)}+\|F_{+,\alpha,h}v_{\alpha,h}\|^{2}_{L^{2}(M; E)}\leq \|F_{-,\alpha,h}v_{\alpha,h}\|^{2}_{L^{2}(M; E)}.  
\end{equation}
Define
\begin{equation*}
    U_{\alpha}:=\{x\in M\mid \mu(x)\leq\alpha\},\ \ s(\alpha):=\sup_{U_{\alpha}}\widetilde{\Phi}.
\end{equation*}
Since $\lambda(h)\rightarrow 0,$ we may take $h$ sufficiently small (depending on $\alpha$) so that
\begin{equation}\label{assmptionsFpm}
    \mathrm{supp}\, F_{-,\alpha,h}\subseteq U_{\alpha}, \ \ F^{2}_{-,\alpha,h}\leq 1,\ \ F^{2}_{+,\alpha,h}\geq\tfrac{1}{2}\alpha^{2}\ \mathrm{on}\ M\backslash U_{\alpha}.
\end{equation}
Therefore,
\begin{equation}\label{F-alphah}
\begin{split}
\|F_{-,\alpha,h}v_{\alpha,h}\|^{2}_{L^{2}(M; E)}&\leq\|v_{\alpha,h}\|^{2}_{L^{2}(U_{\alpha}; E)}
\\
&=\int_{U_{\alpha}}e^{2(1-\alpha)\widetilde{\Phi}/h}|u(h)|^{2}_{E_{x}}dV_{g}
\\
&\leq e^{2(1-\alpha)s(\alpha)/h}.
\end{split}
\end{equation}
Combining \eqref{proofofproposition2.1}, \eqref{assmptionsFpm} and \eqref{F-alphah} shows for some $C_{\alpha}>0$
\begin{equation}\label{eq215}
 \|v_{\alpha,h}\|_{L^{2}(M; E)}+ h\|\nabla v_{\alpha,h}\|_{L^{2}(M;T^{*}M\otimes E)}\leq C_{\alpha}e^{(1-\alpha)s(\alpha)/h}.  
\end{equation}
We now take $w:=e^{\widetilde{\Phi}/h}u(h)=e^{\alpha\widetilde{\Phi}/h}v_{\alpha,h}.$ For a.e. $x\in M,$
\[
h\nabla w=e^{\alpha\widetilde{\Phi}/h}(h\nabla v_{\alpha,h}+\alpha d\widetilde{\Phi}\otimes v_{\alpha,h}).
\]
Hence, if $\epsilon>0$ is given, then \eqref{eq215} gives a $C_{\alpha}'>0$ such that
\[
\|w\|_{L^{2}(M; E)}+h\|\nabla w\|_{L^{2}(M; T^{*}M\otimes E)}\leq C_{\alpha}'\exp\Big(\frac{\alpha\|\widetilde{\Phi}\|_{L^{\infty}(M)}+(1-\alpha)s(\alpha)}{h}\Big)\leq C'_{\alpha}e^{\epsilon/h}.
\]
for $\alpha,h>0$ sufficiently small. This completes the proof.\qed

\subsection{Local WKB construction near a nondegenerate
well} Let $\mu(x)$ be given in \eqref{eigenvalueassumption}. We make the further assumption that there is a neighborhood $U$ of $x_{0}\in M$ such that
\begin{equation}\label{furtherassumption}
 \begin{split}
     &\mathrm{(1)}\ \mu\in C^{\infty}(\overline{U};\mathbb{R})\ \mathrm{is\ a\ simple\ eigenvalue\ for\ each\ }x\in\overline{U},
     \\
     &\mathrm{(2)}\ \mu(x_{0})=0,\ \ d\mu(x_{0})=0,\ \ \mathrm{Hess}_{x_{0}}\mu\gg0,\ \ \mu\big|_{U\backslash \{x_{0}\}}>0.
 \end{split}   
\end{equation}
Shrinking $U$ if necessary, we can construct approximate eigenfunctions of $P_{h}$ on $U$. More precisely, we prove
\begin{prop}\label{wkb}
There exist
\[
a_{m}\in C^{\infty}(U; E),\ \Lambda_{m}\in\mathbb{R},\ m\in\mathbb{N}_{0}
\]
such that, for
\begin{equation}\label{asymptotic}
    a(x;h)\sim\sum_{j=0}^{\infty}h^{j}a_{j}(x),\ \ \Lambda(h)\sim\sum_{j=0}^{\infty}h^{j}\Lambda_{j}
\end{equation}
we have
\begin{equation}\label{approximateeigenfunction}
(P_{h}-h\Lambda(h))\big(e^{-\Phi/h}a\big)=e^{-\Phi/h}r,\ \ r=\mathcal{O}(h^{\infty}).    
\end{equation}
\end{prop}
We refer to \cite[Section 4.4.2]{Zw12} for a discussion on asymptotic sums. We regard $a, a_{m}, \Lambda, \Lambda_{m}, r$ as elements in $S(1).$

We first compute the conjugated operator $e^{\Phi/h}P_{h}e^{-\Phi/h}$, $\Phi\in C^{\infty}(U;\mathbb{R})$. We let $a\in C^{\infty}(U;E)$, $f\in C^{\infty}(U)$, and let $\nabla^{g}$ denote the Levi-Civita connection. Let $(e_{j})_{j=1}^{n}$ be a local orthonormal frame of $TM$. Using the identity 
\begin{equation}\label{bochnerlaplcian}
 \nabla^{*}\nabla=-\sum_{j=1}^{n}\Big(\nabla_{e_{j}}\nabla_{e_{j}}-\nabla_{\nabla^{g}_{e_{j}}e_{j}}\Big),   
\end{equation}
we obtain
\[
\begin{split}
\nabla^{*}\nabla(fa)&=-\sum_{j=1}^{n}\nabla_{e_{j}}\big((e_{j}f)a+f\,\nabla_{e_{j}}a\big)+\sum_{j=1}^{n}\Big(\big((\nabla^{g}_{e_{j}}e_{j})f\big)a+f\,\nabla_{\nabla^{g}_{e_{j}}e_{j}}a\Big) 
\\
&=\sum_{j=1}^{n}\Big(-f\big(\nabla_{e_{j}}\nabla_{e_{j}}-\nabla_{\nabla^{g}_{e_{j}}e_{j}}\big)a-2(e_{j}f)\nabla_{e_{j}}a-\big(e_{j}e_{j}f-(\nabla^{g}_{e_j}e_{j})f\big)a\Big)
\\
&=f\nabla^{*}\nabla a-2\nabla_{\mathrm{grad}_{g}f}a-(\Delta_{g}f)a.
\end{split}
\]
Hence,
\begin{equation}\label{conjugated operator}
e^{\Phi/h}P_{h}e^{-\Phi/h}=\mathcal{B}+h\mathcal{C}+h^{2}\nabla^{*}\nabla,    
\end{equation}
where
\begin{equation}\label{conjugated operator 1}
 \mathcal{B}:=V-|d\Phi|^{2}_{g},\ \ \mathcal{C}:=2\nabla_{X}+\Delta_{g}\Phi,\ \ X:=\mathrm{grad}_{g}\Phi.  
\end{equation}
To achieve \eqref{approximateeigenfunction}, we need to solve the eikonal equation
\begin{equation}\label{Eikonal}
\mathcal{B}a_{0}=0    
\end{equation}
and a sequence of (degenerate) transport equations
\begin{equation}\label{Transport}
\mathcal{B}a_{m}+(\mathcal{C}-\Lambda_{0})a_{m-1}+\nabla^{*}\nabla a_{m-2}-\sum_{l=1}^{m-1}\Lambda_{l}a_{m-1-l}=0,\ \ \ m\geq 1.    
\end{equation}
We first solve \eqref{Eikonal} using
\begin{prop}\label{Eikonal2}
Let $(M, g)$ be a Riemannian manifold and $x_{0}\in M.$ Let $\mu\in C^{\infty}(M; \mathbb{R})$ satisfy
   \[
   \mu(x_{0})=0,\ \ d\mu(x_{0})=0,\ \ \mathrm{Hess}_{x_{0}}\mu\gg0.
   \]
   Then there exists a neighborhood $U\ni x_{0}$ such that the Agmon distance function
   \[
   \begin{split}
    \Phi(x):=d_{\widetilde{g}}(x_{0}, x)
    %\\ & :=\inf_{\gamma}\Big\{\int_{0}^{1}\sqrt{\mu(\gamma(t))}|\dot{\gamma}(t)|_{g}\,\mathrm{d}t\ \Big|\ \gamma(0)=x_{0},\,\gamma(1)=x,\ \gamma\in C^{\infty}([0, 1]; M)\Big\}. 
    \end{split}
\]
is smooth and satisfies
\begin{equation}\label{eikonal3}
 |d\Phi|_{g}^{2}=\mu(x),\ \ \ x\in U.   
\end{equation} 
\end{prop}
\begin{proof}
    1. Define $q: T^{*}M\rightarrow\mathbb{R}$ by
\[
q(x,\xi)=\tfrac{1}{2}|\xi|^{2}_{g}-\tfrac{1}{2}\mu(x).
\]
We first show that $p_{0}:=(x_{0}, 0)$ is a hyperbolic fixed point of $H_{q}.$ That is,
\[
\Spec (dH_{q}(p_{0}))\cap i\mathbb{R}=\emptyset.
\]
We define 
\[
A: T_{x_{0}}M\rightarrow T_{x_{0}}M
\]
by
\[
\mathrm{Hess}_{x_{0}}\mu(u, v)=:g_{x_{0}}(Au, v),\ \ u, v\in T_{x_{0}}M.
\]
By assumption, there exists a $g_{x_{0}}$-orthonormal basis $e_{1},\dots,e_{n}$ such that 
\[
Ae_{j}=a_{j}e_{j},\ \ a_{j}>0.
\]
We choose coordinates $(x_{i})$ centered at $x_{0}$ with $\partial_{i}|_{x_{0}}=e_{i}$, and let $(x,\xi)$ be the induced coordinates on $T^{*}M.$ In these coordinates, we have
\[
H_{q}(x,\xi)=(F(x,\xi), G(x,\xi)),
\]
\[
F^{k}(x,\xi)=g^{kj}(x)\xi_{j},\ \ G^{k}(x,\xi)=\tfrac{1}{2}\partial_{k}\mu(x)-\tfrac{1}{2}\partial_{k}g^{ij}(x)\xi_{i}\xi_{j},
\]
and
\[
\tfrac{\partial F^{k}}{\partial x^{l}}\Big|_{(0,0)}=\partial_{l}g^{kj}(x)\xi_{j}\Big|_{(0,0)}=0,\ \ \tfrac{\partial F^{k}}{\partial\xi_{l}}\Big|_{(0,0)}=g^{kl}(x)\Big|_{(0,0)}=\delta_{kl},
\]
\[
\tfrac{\partial G^{k}}{\partial x^{l}}\Big|_{(0,0)}=\tfrac{1}{2}\partial_{l}\partial_{k}\mu(0)=\tfrac{1}{2}a_{k}\delta_{kl},\ \ \ \ \tfrac{\partial G^{k}}{\partial\xi_{l}}\Big|_{(0,0)}=-\tfrac{1}{2}\partial_{k}g^{ij}(x)(\delta_{il}\xi_{j}+\xi_{i}\delta_{jl})\Big|_{(0,0)}=0.
\]
Thus, $dH_{q}(p_{0})$ has a matrix representation
\[
L=\begin{pmatrix}
    0 & I
    \\
    \tfrac{1}{2}D & 0
\end{pmatrix},
\]
where $D:=\mathrm{diag}(a_{1},\dots,a_{n}).$ We obtain
\[
\Spec (dH_{q}(p_{0}))=\Big\{\pm\sqrt{{a_{j}}/{2}}\mid j=1,\dots,n\Big\}.
\]
Moreover, with respect to the canonical splitting $T_{p_{0}}T^{*}M\cong T_{x_{0}}M\oplus T^{*}_{x_{0}}M,$
\[
dH_{q}(p_{0})(u,\alpha)=(\alpha^{\sharp}, \tfrac{1}{2}(Au)^{\flat}).
\]

2. Let $C:=(\frac{A}{2})^{\frac{1}{2}}$. By the unstable manifold theorem,  $H_{q}$ has a smooth local unstable manifold $\Lambda_{+}\subseteq T^{*}M$ through $p_{0}$, with
\[
T_{p_{0}}\Lambda_{+}\cong\Big\{\big(u, (Cu)^{\flat}\big) \mid u\in T_{x_{0}}M\Big\}.
\]
There exist constants $C_{0}, \lambda>0$ such that for any $\rho\in \Lambda_{\pm}$, $v\in T_{\rho}\Lambda_{+}$
\[
\|d\varphi_{t}(\rho)v\|\leq C_{0}e^{-\lambda |t|}\|v\|,\ \ t\leq 0,
\]
where $\|\cdot\|$ is the Euclidean norm in a neighborhood of $p_{0}=(0,0),$ and $\varphi_{t}:=\exp(tH_{q}).$

Let $\omega$ be the standard symplectic form on $T^{*}M$. Since for $v, w\in T_{\rho}\Lambda_{+},$
\[
\begin{split}
|\omega_{\rho}(v, w)|&=|\omega_{\varphi_{t}(\rho)}(d\varphi_{t}(\rho)v, d\varphi_{t}(\rho)w)|
\\
&\leq C_{\omega}\|d\varphi_{t}(\rho)v\|\|d\varphi_{t}(\rho)w\|
\\
&\leq C_{\omega}C_{0}^{2}e^{-2\lambda|t|}\|v\|\|w\|\xrightarrow[]{ t\rightarrow-\infty} 0,
\end{split}
\]
and $\mathrm{dim}\,\Lambda_{+}=n,$ we conclude that $\Lambda_{+}$ is Lagrangian.
Therefore, there exist a neighborhood $U\ni x_{0}$ and a smooth function $\Phi\in C^{\infty}(U; \mathbb{R})$ with $\Phi(x_{0})=0$ such that
\[
\Lambda_{+}=\{(x, d\Phi(x))\mid x\in U\},
\]
and
\[
\tfrac{1}{2}(|d\Phi(x)|_{g}^{2}-\mu(x))=q(x, d\Phi(x))=q(\varphi_{t}(x,d\Phi(x)))\xrightarrow[]{t\rightarrow-\infty}q(p_{0})=0.
\]

3. It remains to identify $\Phi$ with the Agmon distance.

Suppose $\gamma:[0,1]\rightarrow M$ is a smooth curve from $x_{0}$ to $x.$ Then
\[
\Phi(x)=\int_{0}^{1}d\Phi_{\gamma(t)}(\dot{\gamma}(t))\,\mathrm{d}t\leq\int_{0}^{1}|d\Phi|_{g}\,|\dot{\gamma}|_{g}\,\mathrm{d}t=\ell_{\widetilde{g}}(\gamma).
\]
This shows
\[
\Phi(x)\leq d_{\widetilde{g}}(x_{0}, x).
\]
For the reverse inequality, we take $x\in U$ and define for $t\leq 0$
\[
x(t)=\pi(\varphi_{t}(x,d\Phi(x))),\ \ \pi: T^{*}M\rightarrow M.
\]
Then
\[
\begin{split}
d_{\widetilde{g}}(x_{0}, x)&\leq\int_{-\infty}^{0}\sqrt{\mu(x(t))}|\dot{x}(t)|_{g}\,\mathrm{d}t
\\
&=\int_{-\infty}^{0}|d\Phi(x(t))|_{g}|(d\Phi(x(t)))^{\sharp}|_{g}\,\mathrm{d}t
\\
&=\int_{-\infty}^{0}d\Phi(x(t))(\dot{x}(t))\,\mathrm{d}t
\\
&=\Phi(x).
\end{split}
\]
\end{proof}

We now turn to the transport equations \eqref{Transport}. To diagonalize the equations and construct formal solutions, we make the following definitions. Let $L\rightarrow U$ be the Hermitian line bundle given by
\begin{equation}\label{definelinebundle}
 L:=\bigsqcup\limits_{x\in U}\ker(V|_{E_{x}}-\mu(x)).  
\end{equation}
%Let
%\begin{equation}\label{defineconnection}
 %\nabla^{L}:=(\mathrm{id}_{T^{*}U}\otimes\Pi)\nabla\big|_{L},\   
%\end{equation}
Let $\Pi\in C^{\infty}(U;\mathrm{End}(E)),$ $\Pi_{x}:E_{x}\rightarrow L_{x}$ be the orthogonal projection, and $\mathcal{C}$ be given in \eqref{conjugated operator 1}. We define
\begin{equation}\label{defCL}
\mathcal{C}_{L}:=\Pi\mathcal{C}\big|_{C^{\infty}(U;L)}=2\nabla^{L}_{X}+(\Delta_{g}\Phi),
\end{equation}
where $\nabla^{L}$ is the induced connection on $L$. As in the proof of Proposition \ref{Eikonal2}, we write
\begin{equation}\label{definitionnuj}
\mathrm{Hess}_{x_{0}}\mu(u, v)=g_{x_{0}}(Au, v),\ \ Ae_{j}=a_{j}e_{j},\ \nu_{j}:=\sqrt{\tfrac{a_{j}}{2}}.        
\end{equation}
Let $(x_{i})$ be coordinates centered at $x_{0}$ with $\partial_{i}|_{x_{0}}=e_{i}.$ Then
\[
\Phi(x)=\tfrac{1}{2}\sum_{j=1}^{n}\nu_{j}x_{j}^{2}+\mathcal{O}(|x|^{3}),\ \ X=\sum_{j=1}^{n}\nu_{j}x_{j}\partial_{x_{j}}+\mathcal{O}(|x|^{2}),\ \ \Delta_{g}\Phi=\sum_{j=1}^{n}\nu_{j}+\mathcal{O}(|x|).
\]
Choose a smooth local frame $e_{L}$ of $L$, and write $\nabla^{L}e_{L}=\omega\otimes e_{L}.$ Then
\[
\mathcal{C}_{L}(fe_{L})=\big(2Xf+2\omega (X)f+(\Delta_{g}\Phi)f\big)e_{L}.
\]
It also follows that the homogeneous leading part of $\mathcal{C}_{L}$ is 
\begin{equation}\label{homogeneousleadingformula}
\mathcal{C}_{L,0}(fe_{L}):=\Big(2\sum_{j=1}^{n}\nu_{j}x_{j}\partial_{x_{j}}f+\sum_{j=1}^{n}\nu_{j}f\Big)e_{L}.    
\end{equation}
\emph{Proof of Proposition \ref{wkb}.}
\par 1. We decompose $a_{j}$ as
\[
a_{j}=s_{j}+q_{j},\ \ s_{j}:=\Pi a_{j}\in C^{\infty}(U;L).
\]
Equation \eqref{Eikonal} gives $a_{0}=s_{0}$. We now take
\[
\Lambda_{0}:=\sum_{j=1}^{n}\nu_{j}
\]
and solve 
\begin{equation}\label{scalarhomogeneoustransport}
(\mathcal{C}_{L}-\Lambda_{0})s_{0}=0.    
\end{equation}
We write $\mathcal{C}_{L}=:\mathcal{C}_{L,0}+\mathcal{C}_{L,\geq 1}$ and let $\mathcal{P}_{m}$ be the space of homogeneous polynomials of degree $m$. By
\eqref{homogeneousleadingformula}
    \[
    \mathcal{C}_{L,0}(x^{\alpha}e_{L}(x))=\Big(\sum_{j=1}^{n}(2\alpha_{j}+1)\nu_{j}\Big)x^{\alpha}e_{L}(x),\ \ \alpha\in\mathbb{N}_{0}^{n}.
    \]
    Therefore, $\mathcal{C}_{L,0}-\Lambda_{0}$ is invertible on every homogeneous subspace $\mathcal{P}_{m}e_{L}$ except $\mathbb{C}e_{L}.$

    To construct a formal solution $s_{0}\in\mathbb{C}[[x_{1},\dots,x_{n}]]e_{L}$ to \eqref{scalarhomogeneoustransport}, we set the degree zero term of $s_{0}$ to be $e_{L}$. Suppose that all terms up to degree $N-1$ have been chosen. Since $\mathcal{C}_{L,\geq 1}$ raises the degree by at least one, the degree $N$ term of $s_{0}$ is given by
    \begin{equation}\label{formulafortheprincipalterm}
 s_{0, N}=-(\mathcal{C}_{L,0}-\Lambda_{0})^{-1}\Big[\mathcal{C}_{L,\geq 1}\big(\sum_{0\leq j\leq N-1}s_{0, j}\big)\Big]_{N}.       
    \end{equation}
    2. We now construct the higher-order coefficients of $a(x;h)$ and $\Lambda(h).$ We first project the equation \eqref{Transport} using $\Pi$ to obtain
    \begin{equation}\label{inhomogeneous}
     \begin{split}
(\mathcal{C}_{L}-\Lambda_{0})s_{m-1}&=-\Pi(\mathcal{C}-\Lambda_{0})q_{m-1}-\Pi\nabla^{*}\nabla a_{m-2}+\sum_{l=1}^{m-2}\Lambda_{l}s_{m-1-l}+\Lambda_{m-1}s_{0}
\\
&=:g_{m}+\Lambda_{m-1}s_{0}.
  \end{split}        
    \end{equation}
    We assume all terms on the right-hand side of \eqref{inhomogeneous} are known except for the scalar $\Lambda_{m-1}$. We let
    \[
\widehat{g}_{m}=\Big(\sum_{k=0}^{\infty}G_{m,k}\Big)e_{L},\ \ G_{m,k}\in\mathcal{P}_{k}
\] be the formal Taylor series of $g_{m}.$ Taking the homogeneous degree-zero component of \eqref{inhomogeneous} gives
\[
\Lambda_{m-1}=-G_{m,0}.
\]
We now take $s_{m-1, 0}=0$. Suppose that $s_{m-1,0},\dots,s_{m-1, N-1}$ have been constructed. The degree $N$ term of $s_{m-1}$ is given by
\[
s_{m-1, N}=(\mathcal{C}_{L,0}-\Lambda_{0})^{-1}\Big[g_{m}+\Lambda_{m-1}s_{0}-\mathcal{C}_{L,\geq 1}\big(\sum_{0\leq j\leq N-1}s_{m-1,j}\big)\Big]_{N}.
\]
3. For each $m$, the formal solution $s_{m}\in\mathbb{C}[[x_{1},\dots,x_{n}]]e_{L}$ constructed above can be corrected to an exact smooth solution (see the discussion following \cite[Proposition 3.4]{DiSj99}) provided $U$ satisfies the star-shaped condition with respect to the backward flow of $X=\mathrm{grad}_{g}\Phi:$
\begin{equation}\label{starshape}
 \begin{split}
     &(1)\ \mathrm{For\ every\ }x\in U,\,t\leq 0,\, \exp(t X)(x)\in U\ \mathrm{and\ converges \ to}\ x_{0}\ \mathrm{as}\ t\rightarrow -\infty.
     \\
     &(2)\ \mathrm{For\ every\ compact\ set\ }K\Subset U,\ \{x_{0}\}\cup \{\exp(tX)(x)\mid x\in K, t\leq 0\}\ \mathrm{is\ compact\ in\ }U.
 \end{split}  
\end{equation}
We then project \eqref{Transport} using $1-\Pi$ and obtain 
  \[
  q_{m}:=-(\mathcal{B}|_{(1-\Pi)E})^{-1}(1-\Pi)\Big((\mathcal{C}-\Lambda_{0})a_{m-1}+\nabla^{*}\nabla a_{m-2}-\sum_{l=1}^{m-1}\Lambda_{l}a_{m-1-l}\Big).
  \]
  This completes the proof.
  \qed
\section{Asymptotics of the lowest bands}\label{3}
\subsection{Reduction and geometric properties of the Agmon
distance}
We first reduce $H(k)$ in \eqref{originaloperator} to an operator of the form \eqref{eq2.1}.
\begin{prop}\label{reductionatoddpi}
Suppose $B=(2\ell+1)\pi,\ \ell\in\mathbb{Z}$. Then $H(k)$ is unitarily equivalent to $\widetilde{P}_{h}(k)\in\mathrm{Diff}^{2}(\mathbb{R}^{2}/2\pi\mathbb{Z}^{2};\mathbb{C}^{2})$ given by
\begin{equation}\label{reduced operator}
\widetilde{P}_{h}(k):=h^{2}Q(k)+ V,    
\end{equation}
where
\[
Q(k):=\begin{pmatrix}
    (D_{x_{1}}-k_{1})^{2}+(2D_{x_{2}}-k_{2})^{2} & 0
    \\
    0 & (D_{x_{1}}-k_{1})^{2}+(2D_{x_{2}}+1-k_{2})^{2}
\end{pmatrix}, 
\]
\[
V:=\begin{pmatrix}
 \cos x_{1}  &        \tfrac{1}{2}(1+e^{ix_{2}})
    \\
 \tfrac{1}{2}(1+e^{-ix_{2}})   &    -\cos x_{1}
\end{pmatrix}.
\]
\end{prop}
\begin{proof}
Let
\[
e_{m,n}:=\tfrac{1}{2\pi}e^{imx_{1}+inx_{2}},\ \ (m,n)\in\mathbb{Z}^{2}.
\]
We have
\[
H(k)=h^{2}H_{1}(k)+\tfrac{1}{2}H_{2}(k)+\tfrac{1}{2}H_{3}(k),\ \ \ 
H_{1}(k)e_{m,n}:=\big((m-k_{1})^{2}+(n-k_{2})^{2}\big)e_{m,n},
\]
\[
H_{2}(k)e_{m,n}:=e_{m,n+1}+e_{m,n-1},\ \ H_{3}(k)e_{m,n}:=e^{iB(n-k_{2})}e_{m+1,n}+e^{-iB(n-k_{2})}e_{m-1,n}.
\]

We now conjugate $H(k)$ using the unitary operator $G(k):=e^{iBk_{2}D_{x_{1}}}.$ 
We have
\[
G(k) H_{1}(k)G(k)^{*}=H_{1}(k),\ \ G(k)H_{2}(k)G(k)^{*}=H_{2}(k),
\]
\begin{equation}\label{eq:conjugationgeneralB}
 \begin{split}
G(k)H_{3}(k)G(k)^{*}e_{m,n}&=e^{iB(m+1)k_{2}}e^{iB(n-k_{2})}e^{-iBmk_{2}}e_{m+1,n}\\
&\ \ \ \ \ \ \ \ \ \ \ \ \ \ \ +e^{iB(m-1)k_{2}}e^{-iB(n-k_{2})}e^{-iBmk_{2}}e_{m-1,n}
\\
&=e^{iBn}e_{m+1,n}+e^{-iBn}e_{m-1,n}
\\
&=(-1)^{n}(e_{m+1,n}+e_{m-1,n}).
\end{split}   
\end{equation}

We let $\mathcal{F}: L^{2}(\mathbb{R}^{2}/2\pi\mathbb{Z}^{2})\rightarrow\ell^{2}(\mathbb{Z}^{2})$ be the Fourier series map and define 
\[
\mathcal{E}:\ell^{2}(\mathbb{Z}^{2})\rightarrow\ell^{2}(\mathbb{Z}^{2})\oplus\ell^{2}(\mathbb{Z}^{2}),
\]
\[
\mathcal{E}u:=(a, b),\ \ a_{m,p}:=u_{m,2p},\ \ b_{m,p}:=u_{m, 2p+1}.
\]

If we let
\[
\widetilde{H}_{j}(k):=\mathcal{EF}G(k)H_{j}(k)G(k)^{*}\mathcal{F^{*}E^{*}},
\]
then
\[
\widetilde{H}_{1}(k):(a, b)\mapsto \Big(\big((m-k_{1})^{2}+(2p-k_{2})^{2}\big)a_{m,p}\ ,\ \big((m-k_{1})^{2}+(2p+1-k_{2})^{2}\big)b_{m,p}\Big),
\]
\[
\widetilde{H}_{2}(k):(a, b)\mapsto (b_{m,p-1}+b_{m,p}\ ,\  a_{m,p}+a_{m,p+1}),
\]
\[
\widetilde{H}_{3}(k):(a, b)\mapsto (a_{m-1, p}+a_{m+1,p} \ ,\ -b_{m-1,p}-b_{m+1,p}).
\]
Applying inverse Fourier series to each component gives the claimed operator.
\end{proof}
We replace $V$ in \eqref{reduced operator} by $V+\sqrt{2},$ and take
\begin{equation}\label{metric,connection}
\begin{split}
 g:=dx_{1}^{2}+\tfrac{1}{4}dx_{2}^{2},\ E:=\mathbb{R}^{2}/2\pi\mathbb{Z}^{2}\times\mathbb{C}^{2},\  \langle u, v\rangle_{E_{x}}:=u_{1}\bar{v}_{1}+u_{2}\bar{v}_{2},
 \\
 \nabla^{k}:=d+\theta(k),\ \ \theta(k):=\frac{1}{i}\begin{pmatrix}
  k_{1}dx_{1}+\tfrac{1}{2}k_{2}dx_{2} & 0
  \\
  0 & k_{1}dx_{1}+\tfrac{k_{2}-1}{2}dx_{2}
\end{pmatrix}.
 \end{split}
\end{equation}
%Letting $h:=\gamma^{-\frac{1}{2}}$ be the semiclassical parameter,
We then obtain the semiclassical differential operator
\begin{equation}\label{semiclassicalreduced}
P_{h}(k):=h^{2}(\nabla^{k})^{*}\nabla^{k}+V,\ \ k\in\mathbb{R}^{2}/\mathbb{Z}^{2}.    
\end{equation}
The shifted potential satisfies \eqref{eigenvalueassumption}, and the assumptions \eqref{furtherassumption} hold in a neighborhood of each of its two wells. The two eigenvalues $\mu_{\pm}$ of $V$ are given by
\begin{equation}\label{eigenvaluesofthematrixpotential}
 \mu_{\pm}(x)=\sqrt{2}\pm\sqrt{\cos^{2}x_{1}+\cos^{2}(\tfrac{x_{2}}{2})}.   
\end{equation}
In the notation of \eqref{eigenvalueassumption}, we have
\begin{equation}\label{eigenvalueproperty}
\begin{split}
&\Gamma=\{[(0,0)], [(\pi, 0)]\}, 
\\
 x_{0}\in \Gamma&\ \implies d\mu_{-}(x_{0})=0,\ \mathrm{Hess}_{x_{0}}\,\mu_{-}\gg0,
 \\
 \mu_{-}(x)=\mu_{+}(x)&\ \iff x=(\pm\tfrac{\pi}{2},\pi).
\end{split}
\end{equation}
We now record some useful geometric properties.
\begin{prop}\label{geometricproperties}
 We let $\mu:=\mu_{-}$, $\widetilde{g}:=\mu g$ be the periodic metric on $\mathbb{R}^{2}$. Then
 \begin{enumerate}
     \item [(1)] The Agmon distance from the origin to a lifted well $(m\pi, 2n\pi)$, with $m,n\in\mathbb{Z}$, is given by
     \begin{equation}\label{tunnelingdistance}
     \begin{split}
         d_{\widetilde{g}}((0,0), (m\pi, 2n\pi))=(|m|+|n|)S,\ \ S:=\int_{0}^{\pi}\sqrt{\sqrt{2}-\sqrt{1+\cos^{2}t}}\,\mathrm{dt}   
     \end{split}.
     \end{equation}
     \item [(2)] For any $r\in(0, \pi),$ the distance function $d_{\widetilde{g}}((0,0),\bullet)$ is smooth in a neighborhood $U$ containing $\{(t, 0)\mid 0\leq t\leq r\}.$ Moreover, $U$ is star-shaped in the sense of \eqref{starshape}.
     \item [(3)] Let $\Psi(y):=d_{\widetilde{g}}((0,0), (\frac{\pi}{2},y))+d_{\widetilde{g}}((\pi, 0),(\frac{\pi}{2}, y)).$ Then
     \begin{equation}
    \Psi(0)=S,\ \ \Psi'(0)=0,\ \ \Psi''(0)>0.
     \end{equation}
 \end{enumerate}
\end{prop}
\begin{proof}
   Part (1) is a consequence of the inequality
   \[
   \mu(x_{1}, x_{2})\geq \mu(x_{1}, 0)+\mu(0, x_{2}).
   \]
   Let $\gamma(t)=(x_{1}(t), x_{2}(t)):[0, 1]\rightarrow \mathbb{R}^{2}$ be any smooth curve from $(0,0)$ to $(m\pi, 2n\pi),$ and define
   \[
   F(t):=\int_{0}^{t}\sqrt{\sqrt{2}-\sqrt{1+\cos^{2}s}}\,\mathrm{d}s.
   \]
   Then
   \[
   \begin{split}
   \ell_{\widetilde{g}}(\gamma)&=\int_{0}^{1}\sqrt{\mu(\gamma(t))}\sqrt{\dot{x}_{1}(t)^{2}+\tfrac{1}{4}\dot{x}_{2}(t)^{2}}\,\mathrm{d}t
   \\
   &\geq \int_{0}^{1}\bigg(\sqrt{\mu(x_{1}(t), 0)}|\dot{x}_{1}(t)|+\tfrac{1}{2}\sqrt{\mu(0, x_{2}(t))}|\dot{x}_{2}(t)|\bigg)\,\mathrm{d}t
   \\
   &\geq |F(m\pi)-F(0)|+|F(n\pi)-F(0)|
   \\
   &=(|m|+|n|)S.
   \end{split}
   \]
   The reverse inequality $d_{\widetilde{g}}((0,0), (m\pi, 2n\pi))\leq (|m|+|n|)S$ follows from the fact that if $\gamma$ is a concatenation of a horizontal segment from $(0, 0)$ to $(m\pi, 0)$ and a vertical segment from $(m\pi, 0)$ to $(m\pi, 2n\pi),$ then $\ell_{\widetilde{g}}(\gamma)=(|m|+|n|)S.$

    To prove (2), we fix $\epsilon>0$ sufficiently small. The smoothness of $d_{\widetilde{g}}((0,0),\bullet)$ within the metric ball $B_{\epsilon}:=B_{\widetilde{g}}((0,0), \epsilon)$ was proved in Proposition \ref{Eikonal2}. 

    For $x\notin B_{\epsilon},$ we write
    \[
    d_{\widetilde{g}}((0,0),\bullet)=\epsilon+d_{\widetilde{g}}(\partial B_{\epsilon}, \bullet).
    \]
    Suppose $r, \delta>0$ are fixed and $\epsilon <F(r)<F(r+\delta)<S.$ By part (a), the unit-speed horizontal curve $\gamma:[0, l]\rightarrow\mathbb{R}^{2}$ connecting $\partial B_{\epsilon}\cap \{x_{1}>0, x_{2}=0\}$ and $(r+\delta, 0)$ is a minimizing geodesic. Hence, $\gamma(s)$ is not a focal point to $\partial B_{\epsilon}$ along $\gamma$ for $0<s<l.$ Let $\exp^{\perp}:N\partial B_{\epsilon}\rightarrow \mathbb{R}^{2}$ be the normal exponential map, $\mathcal{K}:=(\gamma(0), [0, l]\gamma'(0))$.  We apply \cite[Chapter~10, Proposition~4.4]{Ca92} to conclude that $d(\exp^{\perp})_{w}$ is invertible for $w\in \mathcal{K}$. 
    
    Since $\exp^{\perp}|_{\mathcal{K}}$ is injective, we obtain a neighborhood $\mathcal{U}$ containing $\mathcal{K}$ such that $\exp^{\perp}|_{\mathcal{U}}$ is a diffeomorphism onto its image. Therefore, it suffices to take 
    \[
    U:=B_{\epsilon}\cup \exp^{\perp}(\mathcal{U}).
    \]

    To prove (3), we use the second variation formula \cite[Theorem 10.22]{Le18} to write
    \[
    \Psi''(0)=\int_{0}^{S}\Big(|D_{t}X^{\perp}|^{2}_{\widetilde{g}}-\mathrm{Rm}(X^{\perp}, \gamma',\gamma', X^{\perp})\Big)\,\mathrm{d}t,
    \]
    where $X^{\perp}$ denotes the normal component of the variation field. Hence, it suffices to show that the curvature term is non-positive. 
    
    We  write $\widetilde{g}=e^{2f}g,\ f:=\frac{1}{2}\log(\mu)$. A computation shows that the Gaussian curvature
    \[
    \begin{split}
    K_{\widetilde{g}}&=-e^{-2f}\Delta_{g}f
    \\
    &=\frac{1-\rho^{2}}{\rho(\sqrt{2}-\rho)^{2}}+\frac{(2\rho-\sqrt{2})(\sin^{2}(2x_{1})+\sin^{2}x_{2})}{8\rho^{3}(\sqrt{2}-\rho)^{3}},
    \end{split}
    \]
    where
    \[
\rho(x_{1}, x_{2}):=\sqrt{\cos^{2}(x_{1})+\cos^{2}(\tfrac{x_{2}}{2})}.
\]
We obtain
\[
K(x_{1}, 0)=-\frac{\sqrt{2}(\rho^{2}-1)}{2\rho^{3}(\sqrt{2}-\rho)}\leq 0.
\]
\end{proof}

\subsection{Isolation of the first two bands}
We justify the existence of a spectral gap of size $h$ between the third and first two bands of $P_{h}(k)$. This is done by comparing $P_{h}(k)$ to two one-well Dirichlet realizations. We let $U_{0}\subseteq\mathbb{R}^{2}/2\pi\mathbb{Z}^{2}\backslash [(\pi, 0)]$, $U_{\pi}\subseteq\mathbb{R}^{2}/2\pi\mathbb{Z}^{2}\backslash [(0,0)]$ be disjoint and contractible open domains satisfying the assumptions \eqref{furtherassumption}, \eqref{starshape}, and consider the Dirichlet realizations $P_{h, U_{0}}(k)$, $P_{h, U_{\pi}}(k)$ with domains $H^{1}_{0}(U_{0};\mathbb{C}^{2})\cap H^{2}(U_{0};\mathbb{C}^{2})$, $H^{1}_{0}(U_{\pi};\mathbb{C}^{2})\cap H^{2}(U_{\pi};\mathbb{C}^{2})$, respectively. 
\begin{prop}\label{simiplicityofonewell}
     Suppose $\Lambda_{0},\nu_{j}$ are given in \eqref{asymptotic}, \eqref{definitionnuj}. If $C_{0}\in(\Lambda_{0}, \Lambda_{0}+2\min_{j}\nu_{j}),$ then there exists $h_{0}>0$ such that for every $0<h<h_{0}$ and every $k\in\mathbb{R}^{2}/\mathbb{Z}^{2}$
      \begin{equation}\label{simplicityofeigenvalue}
          \Spec P_{h, U_{0}}(k)\cap [0,C_{0}h]=\{\lambda(h)\}.
      \end{equation}
      Moreover,  $\lambda(h)$ is simple and has the same asymptotic expansion as $h\Lambda(h)$. The same conclusion holds with $P_{h, U_{0}}(k)$ replaced by $P_{h, U_{\pi}}(k).$
\end{prop}
\begin{proof}
1. Existence of $\lambda(h):$ By \cite[Theorem 7.7.5]{Hormander I}, the normalized approximate eigenfunction on $U_{0}$ has the form
    \begin{equation}
        w(x;h)=h^{-\frac{n}{4}}c(h)e^{-\Phi(x)/h}a(x;h),
    \end{equation}
    where $c(h)=\mathcal{O}(1)$. We now let $\chi\in C^{\infty}_{c}(U;[0,1])$ be a cutoff function near $x_{0}$  and apply \cite[Lemma 6.5]{TaZw23} with $J=1$, $u_{1}:=\frac{\chi w}{\|\chi w\|}$, $\delta=C_{0}h$. The Lemma gives
    \[
    \mathrm{dist}(h\Lambda(h),\Spec P_{h, U})=\mathcal{O}(h^{\infty}).
    \]
    2. Uniqueness of $\lambda(h):$ We first note that the assumptions on $U_{0}$ imply $\Spec P_{h, U_{0}}(k)$ is independent of $k.$ We let 
    \begin{equation}\label{removegauge}
     \mathcal{G}_{k}:=\begin{pmatrix}
         e^{i(k_{1}x_{1}+k_{2}x_{2}/2)} & 0
         \\
         0 & e^{i(k_{1}x_{1}+(k_{2}-1)x_{2}/2)}
     \end{pmatrix},
    \end{equation}
     then $\mathcal{G}_{k}$ is a unitary operator on $L^{2}(U_{0};\mathbb{C}^{2})$ and
    \begin{equation*}
        \mathcal{G}_{k}^{*}P_{h, U_{0}}(k)\mathcal{G}_{k}=-h^{2}\Delta_{g}+\sqrt{2}+\begin{pmatrix}
            \cos x_{1} & \cos(\tfrac{x_{2}}{2})
            \\
            \cos (\tfrac{x_{2}}{2}) & -\cos x_{1}
        \end{pmatrix}.
    \end{equation*}
    Since $\mu_{-}<\mu_{+}$ on $\overline{U}_{0}$, there exists $T\in C^{\infty}(\overline{U}_{0}; U(2))$ such that
    \begin{equation*}
    \begin{split}
        \widetilde{P}_{h}:=T^{*}\mathcal{G}_{k}^{*}P_{h, U_{0}}(k)\mathcal{G}_{k}
        T&=P_{h}^{\mathrm{diag}}+R_{h},
        \\
        P_{h}^{\mathrm{diag}}:=-h^{2}\Delta_{g}+\begin{pmatrix}
            \mu_{-} & 0
            \\
            0 & \mu_{+}
        \end{pmatrix}&,\ R_{h}:=T^{*}[-h^{2}\Delta_{g}, T]\in h\mathrm{Diff}^{1}_{h}(U_{0};\mathbb{C}^{2}).
        \end{split}
    \end{equation*}
    By \cite[Theorem 4.23]{DiSj99}, there exists a contour $\Gamma_{h}$ enclosing $[0, C_{0}h]$ such that
    \begin{equation}\label{contour}
        \mathrm{dist}(\Gamma_{h}, \Spec P_{h}^{\mathrm{diag}})\geq ch
    \end{equation}
    for some $c>0$, and the corresponding Riesz projector of $P_{h}^{\mathrm{diag}}$ has rank one. 

    We now consider the family
    \begin{equation*}
        P_{h}(t):=P_{h}^{\mathrm{diag}}+t R_{h},\ 0\leq t\leq 1,
    \end{equation*}
    with the common domain $H^{2}(U_{0};\mathbb{C}^{2})\cap H^{1}_{0}(U_{0};\mathbb{C}^{2}).$ Let $z\in\Gamma_{h}$, $f\in L^{2}(U_{0};\mathbb{C}^{2})$ and set $u:=(P_{h}^{\mathrm{diag}}-z)^{-1}f.$ \eqref{contour} gives
    \begin{equation}\label{simpleeigenvalue1}
     \|u\|_{L^{2}}\lesssim h^{-1}\|f\|_{L^{2}}.  
    \end{equation}
    Since $\mu_{\pm}>0,$
    \begin{equation}\label{simpleeigenvalue2}
    \begin{split}
   \|hdu\|^{2}_{L^{2}}\leq \langle P^{\mathrm{diag}}_{h}u, u\rangle&\leq |z|\|u\|^{2}_{L^{2}}+\|f\|_{L^{2}}\|u\|_{L^{2}}
   %\\ &
   \lesssim h^{-1}\|f\|^{2}_{L^{2}}.
   \end{split}
   \end{equation}
   Hence,
   \begin{equation}\label{simpleeigenvalue3}
   \begin{split}
 \|R_{h}u\|_{L^{2}}&\lesssim h\|h du\|_{L^{2}}+h^{2}\|u\|_{L^{2}}
 \lesssim h^{\frac{1}{2}}\|f\|_{L^{2}}.
 \end{split}
 \end{equation}
 For sufficiently small $h,$    
  \begin{equation*}
      P_{h}(t)-z=\Big(I+tR_{h}(P^{\mathrm{diag}}_{h}-z)^{-1}\Big)(P_{h}^{\mathrm{diag}}-z)
  \end{equation*} 
  is invertible for $t\in [0, 1],$ $z\in\Gamma_{h}.$ We denote
  \[
  \Pi_{h}(t):=\frac{1}{2\pi i}\int_{\Gamma_{h}}(z-P_{h}(t))^{-1}\,dz
  \]
  and apply \cite[Lemma I.4.10]{Ka95} to obtain
  \[
  \mathrm{rank}\,\Pi_{h}(0)=\mathrm{rank}\,\Pi_{h}(1)=1.
  \]
  This concludes the proof of the proposition. \end{proof}

\par To compare the spectrum of the global operator $P_{h}(k)$ to the Dirichlet realizations, we prove a simple generalization of \cite[Lemma 6.5]{TaZw23}.
\begin{lemm}\label{comparison}
Let $A\geq 0$ be a self-adjoint operator on a Hilbert space $\mathcal{H}$, let $0\leq a< b,$ and suppose $w_{1},\dots,w_{J}\in \mathcal{D}(A)$ and $z_{j}\in [0, a]$ satisfy 
\begin{equation}\label{almostorthogonality}
 |\langle w_{i}, w_{j}\rangle-\delta_{ij}|\leq\eta.   
\end{equation}
If 
\begin{equation}\label{comparisonassumption}
J\eta+\frac{1}{(b-a)^{2}}\sum_{j=1}^{J}\|(A-z_{j})w_{j}\|^{2}<1,    
\end{equation}
then
\[
\mathrm{rank}\,\mathrm{\bf{1}}_{[0, b]}(A)\geq J.
\]
\end{lemm}
\begin{proof}
    We set $\Pi:=\mathrm{\bf{1}}_{[0, b]}(A)$, $Q:=1-\Pi.$ The Spectral Theorem gives
    \[
    \|(A-z_{j})w_{j}\|^{2}\geq \|(A-z_{j})Qw_{j}\|^{2}\geq (b-a)^{2}\|Qw_{j}\|^{2}.
    \]
    Therefore,
    \begin{equation}\label{comparisoneq1}
        \sum_{j=1}^{J}\|Qw_{j}\|^{2}\leq\frac{1}{(b-a)^{2}}\sum_{j=1}^{J}\|(A-z_{j})w_{j}\|^{2}.
    \end{equation}
    Suppose $c=(c_{1},\dots,c_{J})\in\mathbb{C}^{J}$. Assumption \eqref{almostorthogonality} gives
    \begin{equation}\label{comparisoneq2}
\begin{split}
\Big\|\sum_{j=1}^{J}c_{j}w_{j}\Big\|^{2}&=\sum_{i,j=1}^{J}c_{i}\overline{c_{j}}\langle w_{i}, w_{j}\rangle  
\\
&\geq |c|^{2}-\eta\Big(\sum_{j=1}^{J}|c_{j}|\Big)^{2}
\\
&\geq (1-J\eta)|c|^{2}.
\end{split}
    \end{equation}
    \eqref{comparisoneq1}, \eqref{comparisoneq2} then give
  \[
  \begin{split}
  \Big\|\sum_{j=1}^{J}c_{j}\Pi w_{j}\Big\|^{2}&\geq\Big\|\sum_{j=1}^{J}c_{j}w_{j}\Big\|^{2}-|c|^{2}\sum_{j=1}^{J}\|Qw_{j}\|^{2}
  \\
  &\geq \bigg(1-J\eta-\frac{1}{(b-a)^{2}}\sum_{j=1}^{J}\|(A-z_{j})w_{j}\|^{2}\bigg)|c|^{2}.
  \end{split}
  \]
Hence, 
$ 
\sum_{j=1}^{J}c_{j}\Pi w_{j}=0 $ implies $c=0.
$
\end{proof}

\begin{prop}\label{globaloperatorspectralgap}
Let $C_{0}\in(\Lambda_{0}, \Lambda_{0}+2\min_{j}\nu_{j})$. Then there exists $h_{0}>0$ such that for every $0<h<h_{0}$ and every $k\in\mathbb{R}^{2}/\mathbb{Z}^{2},$
\begin{equation}
    \Spec P_{h}(k)\cap[0, C_{0}h]=\{\lambda_{1}(h,k),\lambda_{2}(h,k)\},\ \mathrm{rank}\,\mathrm{\bf{1}}_{[0, C_{0}h]}(P_{h}(k))=2.
\end{equation}
\end{prop}
\begin{proof}
 1. We first establish the lower bound
 \begin{equation}\label{lowerbound}
 \mathrm{rank}\,{\bf{1}}_{[0, C_{0}h]}(P_{h}(k))\geq 2.
 \end{equation}
 Let $\sigma\in\{0,\pi\}$, $C_{\pm}>0$ satisfy $0<\Lambda_{0}<C_{-}<C_{0}<C_{+}<\Lambda_{0}+2\min_{j}\nu_{j}.$ By Proposition \ref{simiplicityofonewell}, there exists a normalized eigenfunction $\varphi_{\sigma}$ such that
 \begin{equation*}
     P_{h, U_{\sigma}}(k)\varphi_{\sigma}=\mu_{\sigma}\varphi_{\sigma},\ \ 0\leq\mu_{\sigma}\leq C_{-}h,
 \end{equation*}
and
\begin{equation*}
    \mathrm{rank}\,\mathrm{\bf{1}}_{[0, C_{+}h]}(P_{h, U_{\sigma}}(k))=1.
\end{equation*}
 
 We choose cut-off functions $\chi_{\sigma}\in C^{\infty}(\mathbb{R}^{2}/2\pi\mathbb{Z}^{2};[0, 1])$ such that
 \[
 \chi_{\sigma}\equiv 1\ \mathrm{near}\ x_{\sigma},\ \ \mathrm{supp}\,\chi_{\sigma}\subseteq U_{\sigma},
 \]
and define $v_{\sigma}\in C^{\infty}(\mathbb{R}^{2}/2\pi\mathbb{Z}^{2};\mathbb{C}^{2})$ by
\[
v_{\sigma}:=\chi_{\sigma}\varphi_{\sigma}.
\]
Proposition \ref{exponentialdecay} (the proof applies to compact manifolds with boundary after imposing the Dirichlet boundary condition) applied to $\varphi_{\sigma}$ gives
\begin{equation*}\label{vsigma1}
\begin{split}
    \|v_{\sigma}\|_{L^{2}}^{2}&=1+\mathcal{O}(e^{-c/h}),
    \\
    (P_{h}(k)-\mu_{\sigma})v_{\sigma}&=[P_{h}(k), \chi_{\sigma}]\varphi_{\sigma}=\mathcal{O}_{L^{2}}(e^{-c/h})
    \end{split}
\end{equation*}
for some $c>0,$ and the disjointness of the supports gives 
$\langle v_{0}, v_{\pi}\rangle=0.$
Hence, \eqref{lowerbound} follows from  Lemma \ref{comparison} with 
\begin{equation*}\label{comparison1}
 A=P_{h}(k),\ \ a=C_{-}h,\ \  b=C_{0}h,\ \  J=2.   
\end{equation*}
\par 2. We now prove the reverse inequality
\begin{equation*}\label{upperbound}
\mathrm{rank}\,{\bf{1}}_{[0, C_{0}h]}(P_{h}(k))\leq 2.   
\end{equation*}
Let $P_{D}(k):=P_{h, U_{0}}(k)\oplus P_{h, U_{\pi}}(k).$ By Proposition \ref{simiplicityofonewell},
\begin{equation*}\label{rankofdirectsum}
\mathrm{rank}\,{\bf{1}}_{[0, C_{+}h]}(P_{D}(k))=2.    
\end{equation*}
Suppose, to the contrary, that
\begin{equation*}
    \mathrm{rank}\,{\bf{1}}_{[0, C_{0}h]}(P_{h}(k))\geq 3.
\end{equation*}
Then there exist orthonormal eigenfunctions $u_{1}, u_{2}, u_{3}\in C^{\infty}(\mathbb{R}^{2}/2\pi\mathbb{Z}^{2})$ satisfying
\begin{equation*}
    P_{h}(k)u_{j}=\lambda_{j}u_{j},\ \ 0\leq\lambda_{j}\leq C_{0}h.
\end{equation*}
We set $w_{j}:=(\chi_{0}u_{j}, \chi_{\pi}u_{j}).$ Applying Proposition \ref{exponentialdecay} to $u_{j}$ gives
\begin{equation*}
 \begin{split}
     \langle w_{i}, w_{j}\rangle=\delta_{ij}-\langle(1-\chi_{0}^{2}-\chi_{\pi}^{2})u_{i}, u_{j}\rangle=\delta_{ij}+\mathcal{O}(e^{-c/h}),
     \\
     (P_{D}(k)-\lambda_{j})w_{j}=([P_{h}(k),\chi_{0}]u_{j}, [P_{h}(k),\chi_{\pi}]u_{j})=\mathcal{O}_{L^{2}}(e^{-c/h})
 \end{split}   
\end{equation*}
for some $c>0.$ Applying Lemma \ref{comparison} with
\[
A=P_{D}(k),\ \ a=C_{0}h,\ \ b=C_{+}h,\ \ J=3
\]
then completes the proof of the proposition.
\end{proof}
\subsection{Asymptotics of the effective matrix}
In this subsection we adapt the arguments from \cite{Ou87} to prove Proposition \ref{asymptoticeffecitivematrix}. Taking the eigenvalues of \eqref{ematrx} then gives
\begin{equation}\label{globalasymptotics1}
\begin{split}
E_{1}(k)&=-\sqrt{2}+\lambda_{h}-2\sqrt{|\tau_{h}|^{2}\cos^{2}(\pi k_{1})+s_{h}^{2}\cos^{2}(\pi k_{2})}+\widetilde{\mathcal{O}}(e^{-2S/h}),
\\
E_{2}(k)&=-\sqrt{2}+\lambda_{h}+2\sqrt{|\tau_{h}|^{2}\cos^{2}(\pi k_{1})+s_{h}^{2}\cos^{2}(\pi k_{2})}+\widetilde{\mathcal{O}}(e^{-2S/h}).
\end{split}
\end{equation}

We use localized one-well Dirichlet eigenfunctions of $P_{h}(k)$ on $(\mathbb{R}^{2}, g)$ to build an orthonormal basis of the eigenspace associated with the first two bands of $P_{h}(k)$. The Dirichlet domains need to be sufficiently large to capture the exponentially small terms in \eqref{exponentiallysmallentries}.

 Let $x_{0}:=(0,0), \ x_{\pi}:=(\pi, 0)$. Fix $0<\delta<S$, $0<\eta\ll S,$ where $S$ is the tunneling distance defined by \eqref{tunnelingdistance}. Let $\Omega_{0}\subseteq\mathbb{R}^{2}$ be a smooth regularization of 
\begin{equation}\label{dirichletdomainoncover}
B_{\widetilde{g}}(x_{0},S+\delta)\backslash\bigcup_{y\in\mathcal{N}_{0}}\overline{B_{\widetilde{g}}(y, \eta)},\ \ \mathcal{N}_{0}:=\{(\pm\pi, 0),(0,\pm 2\pi)\}.    
\end{equation}
By Proposition \ref{geometricproperties}, $\Omega_{0}$ contains precisely one lifted well $x_{0}.$ We define $\Omega_{\pi}:=\Omega_{0}+x_{\pi}$. Let $k_{0}=(\frac{1}{2},\frac{1}{2})$ and
\begin{equation}\label{dirichletrealizationsoncover}
 P_{h, \Omega_{\sigma}}(k_{0}),\ \ \sigma\in\{0,\pi\},   
\end{equation}
be the corresponding Dirichlet realizations. The proof of Proposition \ref{simiplicityofonewell} and the fact that there exists a unitary map $\widetilde{S}_{0}:L^{2}(\Omega_{0};\mathbb{C}^{2})\rightarrow L^{2}(\Omega_{\pi};\mathbb{C}^{2}),$
\begin{equation*}
    (\widetilde{S}_{0}u)(x_{1}, x_{2}):=Au(x_{1}-\pi,-x_{2}),\ \ A:=\begin{pmatrix}
     0 & 1
     \\
     1 & 0
 \end{pmatrix},
\end{equation*}
such that
\begin{equation}\label{unitarymap}
\begin{split}
 \widetilde{S}_{0}P_{h, \Omega_{0}}(k_{0})\widetilde{S}^{*}_{0}=P_{h,\Omega_{\pi}}(k_{0}),  
 \end{split}
\end{equation}
imply each $P_{h,\Omega_{\sigma}}(k_{0})$ has exactly one eigenvalue $\lambda_{h}\in [0, C_{-}h]$. Let $\phi_{\sigma, h}$ be the corresponding normalized eigenfunction
\begin{equation}\label{dirichleteigenfunctiononcover}
 P_{h,\Omega_{\sigma}}(k_{0})\phi_{\sigma, h}=\lambda_{h}\phi_{\sigma, h},\ \ \phi_{\pi, h}=\widetilde{S}_{0}\phi_{0, h}.   
\end{equation}
We now choose cutoff functions $\chi_{\sigma}\in C^{\infty}_{c}(\Omega_{\sigma};[0, 1])$ such that
\begin{equation}\label{cutoffoncover}
\chi_{\pi}=\widetilde{S}_{0}\chi_{0}\widetilde{S}^{*}_{0}, \ \ \chi_{\sigma}\equiv 1\ \mathrm{near}\ x_{\sigma},\ \ \ \mathrm{supp}\,d\chi_{\sigma}\subseteq\{x\in\Omega_{\sigma}\mid \mathrm{dist}(x, \partial\Omega_{\sigma})<\eta\}.
\end{equation}
Define $f_{\sigma, h}, r_{\sigma, h}\in C^{\infty}_{c}(\mathbb{R}^{2};\mathbb{C}^{2})$ by
\begin{equation}\label{fixedaeigenfunction}
    f_{\sigma, h}:=\chi_{\sigma}\phi_{\sigma, h},\ \ \ r_{\sigma, h}:=(P_{h, \Omega_{\sigma}}(k_{0})-\lambda_{h})f_{\sigma, h}=[P_{h,\Omega_{\sigma}}(k_{0}),\chi_{\sigma}]\phi_{\sigma,h}.
    \end{equation}
Let $\kappa:=k-k_{0},$ $\gamma_{m,n}:=(2\pi m, 2\pi n).$ We set
\begin{equation}\label{blochtransform}
    \mathcal{G}_{\kappa}(x):=e^{i(\kappa_{1}x_{1}+\tfrac{\kappa_{2}}{2}x_{2})},\ \ T_{m,n}u(x):=u(x-\gamma_{m,n})
\end{equation}
and define the approximate eigenfunctions
\begin{equation}\label{aeigenfucntion}
    \Psi_{\sigma, h}(\kappa):=\sum_{m,n\in\mathbb{Z}}T_{m,n}(\mathcal{G}_{\kappa}f_{\sigma, h}).
\end{equation}
If
\begin{equation*}
    \mathcal{R}_{\sigma, h}(\kappa):=\sum_{m,n\in\mathbb{Z}}T_{m,n}(\mathcal{G}_{\kappa}r_{\sigma, h}),
\end{equation*}
we then have the identity
\begin{equation}\label{periodicidentity}
    (P_{h}(k_{0}+\kappa)-\lambda_{h})\Psi_{\sigma, h}(\kappa)=\mathcal{R}_{\sigma, h}(\kappa).
\end{equation}
    
To simplify notation, we use
\[
B=\widetilde{\mathcal{O}}(e^{-A/h})\iff \forall\epsilon>0, \ B=\mathcal{O}(e^{-(A-C\eta-\epsilon)/h})\ \mathrm{for\ some\ }C>0\ \mathrm{independent\ of\ } \epsilon\ \mathrm{and}\ \eta.
\]

%We then have the following estimates
%\begin{lemm}\label{interactionmtxestimates}
%Let $f_{\sigma, h}, r_{\sigma, h}, \Psi_{\sigma, h}(\kappa), \mathcal{R}_{\sigma}(\kappa)$ be defined above. Then for every $s\in\mathbb{R},$ we have
%\end{lemm}
\begin{prop}\label{asymptoticeffecitivematrix}
Let $\kappa\in[-\frac{1}{2},\frac{1}{2}]^{2}, \,k=k_{0}+\kappa.$ Suppose $C_{0}>0$ satisfies the assumption in Proposition \ref{simiplicityofonewell}. Then for each $k$, there is an orthonormal basis $(e_{0,h}(k), e_{\pi, h}(k))$ of 
\begin{equation*}
 \mathcal{F}_{h}(k):=\mathrm{Ran}\,\mathrm{\bf{1}}_{[0, C_{0}h]}(P_{h}(k))   
\end{equation*}
in which the operator $P_{h}(k)\big|_{\mathcal{F}_{h}(k)}$ has the matrix representation
\begin{equation}\label{ematrx}
    M_{h}(k)=\lambda_{h}I+\begin{pmatrix}
        2s_{h}\cos(\pi k_{2}) & \tau_{h}(1+e^{-2\pi ik_{1}})
        \\
        \overline{\tau_{h}}(1+e^{2\pi ik_{1}}) & -2s_{h}\cos(\pi k_{2})
    \end{pmatrix}+R_{h}(k).
\end{equation}
Here $\lambda_{h}$ is given in \eqref{dirichleteigenfunctiononcover}, $s_{h}\in\mathbb{R},$ $\tau_{h}\in i\mathbb{R}$, and $R_{h}(k)$ is Hermitian. Moreover,
\begin{equation}\label{exponentiallysmallentries}
    |s_{h}|+|\tau_{h}|=\widetilde{\mathcal{O}}(e^{-S/h}),\ \ \sup_{k}\|R_{h}(k)\|=\widetilde{\mathcal{O}}(e^{-2S/h}).
\end{equation}
\end{prop}
\begin{proof}
    1. We construct a basis of $\mathcal{F}_{h}(k_{0}+\kappa)$ by projecting the approximate eigenfunctions $\Psi_{\sigma, h}(\kappa)$ defined in \eqref{aeigenfucntion}.  Let $G_{\Psi}(\kappa)$ be the Gram matrix of $(\Psi_{0, h}(\kappa), \Psi_{\pi, h}(\kappa))$. Since
    \begin{equation*}
   T_{m,n}\mathcal{G}_{\kappa}=e^{-i(2\pi m\kappa_{1}+\pi n\kappa_{2})}\mathcal{G}_{\kappa}T_{m,n},     
    \end{equation*}
     we have
    \begin{equation*}
    \begin{split}
        (G_{\Psi}(\kappa))_{\alpha\beta}&:=\langle\Psi_{\alpha, h}(\kappa), \Psi_{\beta,h}(\kappa) \rangle_{L^{2}(\mathbb{R}^{2}/2\pi\mathbb{Z}^{2};\mathbb{C}^{2})}
        \\
        &\ =\Big\langle\sum_{m,n\in\mathbb{Z}}T_{m,n}(\mathcal{G}_{\kappa}f_{\alpha, h}), \sum_{m,n\in\mathbb{Z}}T_{m,n}(\mathcal{G}_{\kappa}f_{\beta,h}) \Big\rangle_{L^{2}(\mathbb{R}^{2}/2\pi\mathbb{Z}^{2};\mathbb{C}^{2})}
        \\
        & \ =\sum_{m,n\in\mathbb{Z}}e^{i(2\pi m\kappa_{1}+\pi n\kappa_{2})}\langle f_{\alpha, h}, T_{m,n}f_{\beta,h}\rangle_{L^{2}(\mathbb{R}^{2};\mathbb{C}^{2})}.
        \end{split}
    \end{equation*}
  Now Proposition \ref{exponentialdecay} together with the elliptic estimate \cite[Proposition 6.39]{Hi25} gives
  \begin{equation}\label{pointwisedecay}
      |\partial^{\alpha}\phi_{\sigma, h}(x)|=\widetilde{\mathcal{O}}(e^{-d_{\widetilde{g}}(x_{\sigma}, x)/h})\ \ \forall\alpha\in\mathbb{N}_{0}^{2},\ x\in \Omega_{\sigma}.
  \end{equation}
  Hence,
  \begin{equation}\label{pointwiseconsequence}
  \begin{split}
      \|f_{\sigma, h}\|^{2}_{L^{2}(\mathbb{R}^{2};\mathbb{C}^{2})}=1+\widetilde{\mathcal{O}}(e^{-2S/h}),&\ \ \ |\langle f_{\alpha, h}, T_{mn}f_{\beta,h}\rangle_{L^{2}(\mathbb{R}^{2};\mathbb{C}^{2})}|=\widetilde{\mathcal{O}}(e^{-d_{\widetilde{g}}(x_{\alpha}, x_{\beta}+\gamma_{mn})/h}),
      \\
      G_{\Psi}(\kappa)&=I+\widetilde{\mathcal{O}}(e^{-S/h}).
      \end{split}
  \end{equation}

 We put $\Pi_{h}(\kappa):=\mathrm{\bf{1}}_{[0, C_{0}h]}(P_{h}(k_{0}+\kappa))$, $Q_{h}(\kappa):=1-\Pi_{h}(\kappa)$ and define
    \begin{equation}\label{projectedaeigenfunctions}
    v_{\sigma, h}(\kappa):=\Pi_{h}(\kappa)\Psi_{\sigma, h}(\kappa),\ \ q_{\sigma, h}(\kappa):=Q_{h}(\kappa)\Psi_{\sigma, h}(\kappa).    
    \end{equation}
By Proposition \ref{globaloperatorspectralgap}, \eqref{periodicidentity}, and \eqref{pointwisedecay}, we have
\begin{equation}\label{qestimate}
    \begin{split}
        \widetilde{\mathcal{O}}(e^{-S/h})&=\|\mathcal{R}_{\sigma, h}(\kappa)\|_{L^{2}}
        \geq\|Q_{h}(\kappa)\mathcal{R}_{\sigma, h}(\kappa)\|_{L^{2}}
        \\
        &=\|(P_{h}(k_{0}+\kappa)-\lambda_{h})q_{\sigma, h}(\kappa)\|_{L^{2}}
       \\ & =\|(Q_{h}(\kappa)P_{h}(k_{0}+\kappa)-\lambda_{h})q_{\sigma, h}(\kappa)\|_{L^{2}}
        \\ & \gtrsim h\|q_{\sigma, h}(\kappa)\|_{L^{2}}.
    \end{split}
\end{equation}
Hence, if we let $G_{v}(\kappa)$, $G_{q}(\kappa)$ denote the Gram matrix of $(v_{0, h}(\kappa), v_{\pi, h}(\kappa)),$ $(q_{0, h}(\kappa), q_{\pi, h}(\kappa)),$ respectively, then
\begin{equation*}
    G_{v}(\kappa)=G_{\Psi}(\kappa)-G_{q}(\kappa),\ \ G_{q}(\kappa)=\widetilde{\mathcal{O}}(e^{-2S/h}).
\end{equation*}
Together with \eqref{pointwiseconsequence}, we obtain
\begin{equation*}
    G_{v}(\kappa)=I+\widetilde{\mathcal{O}}(e^{-S/h}),
\end{equation*}
$(v_{0, h}(\kappa), v_{\pi, h}(\kappa))$ is a basis of $\mathcal{F}_{h}(k)$ for sufficiently small $h.$
\par 2. We define 
\[
B_{v}(\kappa)_{\alpha\beta}:=\langle P_{h}(k_{0}+\kappa)v_{\alpha, h}(\kappa), v_{\beta, h}(\kappa)\rangle_{L^{2}},\ \ \ W_{h}(\kappa)_{\alpha\beta}:=\langle\mathcal{R}_{\alpha, h}(\kappa), \Psi_{\beta,h}(\kappa)\rangle_{L^{2}}=\widetilde{\mathcal{O}}(e^{-S/h}).
\]
Using \eqref{periodicidentity} and \eqref{qestimate}, we obtain
\begin{equation*}
    \begin{split}
        B_{v}(\kappa)_{\alpha\beta}&=\langle P_{h}(k_{0}+\kappa)\Psi_{\alpha, h}(\kappa), \Psi_{\beta, h}(\kappa)\rangle_{L^{2}}-\langle P_{h}(k_{0}+\kappa)q_{\alpha, h}(\kappa), q_{\beta, h}(\kappa)\rangle_{L^{2}}
        \\
        &=\lambda_{h}\langle v_{\alpha, h}(\kappa),v_{\beta,h}(\kappa)\rangle_{L^{2}}+\langle\mathcal{R}_{\alpha, h}(\kappa),\Psi_{\beta,h}(\kappa)\rangle_{L^{2}}
        \\
        &\hspace{43mm}-\langle(P_{h}(k_{0}+\kappa)-\lambda_{h})q_{\alpha, h}(\kappa), q_{\beta, h}(\kappa)\rangle_{L^{2}}
        \\
        &=\lambda_{h}G_{v}(\kappa)_{\alpha\beta}+W_{h}(\kappa)_{\alpha\beta}+\widetilde{\mathcal{O}}(e^{-2S/h}).
    \end{split}
\end{equation*}
We now define the orthonormal basis of $\mathcal{F}_{h}(k_{0}+\kappa)$ by
\begin{equation}\label{orthonormalize}
 \begin{pmatrix}
     e_{0,h}(\kappa)
     \\
     e_{\pi, h}(\kappa)
 \end{pmatrix}:=G_{v}(\kappa)^{-\frac{1}{2}}\begin{pmatrix}
     v_{0, h}(\kappa)
     \\
     v_{\pi, h}(\kappa)
 \end{pmatrix}.
\end{equation}
Then
\begin{equation*}
\begin{split}
    M_{h}(k_{0}+\kappa)&=G_{v}^{-\frac{1}{2}}(\kappa)B_{v}(\kappa)G^{-\frac{1}{2}}_{v}(\kappa)
    \\
    &=\lambda_{h}I+(I+\widetilde{\mathcal{O}}(e^{-S/h}))(W_{h}(\kappa)+\widetilde{\mathcal{O}}(e^{-2S/h}))(I+\widetilde{\mathcal{O}}(e^{-S/h}))
    \\
    &=\lambda_{h}I+W_{h}(\kappa)+\widetilde{\mathcal{O}}(e^{-2S/h}).
    \end{split}   
\end{equation*}
\ \ 3. We now evaluate the interaction matrix $W_{h}(\kappa)$ to derive \eqref{ematrx}. We write
\begin{equation*}
    W_{h}(\kappa)_{\alpha\beta}=\sum_{m,n\in\mathbb{Z}}e^{i(2\pi m\kappa_{1}+\pi n\kappa_{2})}w_{\alpha\beta,h}(m,n),
\end{equation*}
where
\begin{equation}\label{interactionmatrixcoefficients}
    w_{\alpha\beta,h}(m,n):=\langle r_{\alpha, h}, T_{m,n}f_{\beta,h}\rangle_{L^{2}(\mathbb{R}^{2};\mathbb{C}^{2})},\ \ \ w_{\beta\alpha, h}(-m,-n)=\overline{w_{\alpha\beta,h}(m,n)}.
\end{equation}
By \eqref{pointwisedecay}, we have
\begin{equation*}
    |w_{\alpha\beta,h}(m,n)|=\widetilde{\mathcal{O}}(e^{-d_{\widetilde{g}}(x_{\alpha}, x_{\beta}+\gamma_{m,n})/h}).
\end{equation*}
Moreover, since $\mathrm{supp}\,r_{\sigma, h}\subseteq\{x\in\Omega_{\sigma}\mid d_{\widetilde{g}}(x, x_{\sigma})\geq S-2\eta\}$ and $r_{\sigma, h}, f_{\sigma, h}$ contain the same Dirichlet eigenfunction, we have
\begin{equation*}
    |w_{\sigma\sigma,h}(0,0)|=\widetilde{\mathcal{O}}(e^{-2S/h}).
\end{equation*}
Hence,
\begin{equation*}
 W_{h}(\kappa)=\begin{pmatrix}
     w_{00,h}(0,1)e^{i\pi\kappa_{2}}+\overline{w_{00,h}(0,1)}e^{-i\pi\kappa_{2}} & w_{0\pi,h}(0,0)+w_{0\pi, h}(-1, 0)e^{-2\pi i\kappa_{1}}
     \\
     \overline{w_{0\pi, h}(0,0)}+\overline{w_{0\pi,h}(-1, 0)}e^{2\pi i\kappa_{1}} & w_{\pi\pi,h}(0, 1)e^{i\pi\kappa_{2}}+\overline{w_{\pi\pi, h}(0, 1)}e^{-i\pi\kappa_{2}}
 \end{pmatrix}+\widetilde{\mathcal{O}}(e^{-2S/h}). 
\end{equation*}
From \eqref{unitarymap}-\eqref{blochtransform}, we get
\begin{equation*}
\begin{split}
    f_{\pi, h}=\widetilde{S}_{0}f_{0, h},\ \ r_{\pi, h}=\widetilde{S}_{0}r_{0, h},\ \ \widetilde{S}_{0}^{2}=T_{1, 0},\ \ \widetilde{S}_{0}T_{m,n}=T_{m,-n}\widetilde{S}_{0}.
    \\
    w_{\pi\pi, h}(0, 1)=\overline{w_{00, h}(0, 1)},\ \ \ w_{0\pi, h}(-1, 0)=\overline{w_{0\pi,h}(0,0)}.
    \end{split}
\end{equation*}
We now show that both $w_{00, h}(0, 1)$ and $w_{0\pi, h}(0,0)$ are purely imaginary. Let $\widetilde{S}_{1}$ be the antiunitary operator
\[
(\widetilde{S}_{1}u)(x):=\begin{pmatrix}
e^{i(x_1+x_2/2)}\overline{u_1(x)}\\
e^{i(x_1-x_2/2)}\overline{u_2(x)}
\end{pmatrix}.
\]
A direct calculation shows that
\[
[\widetilde{S}_{1}, P_{h}(k_{0})]=0,\ \ \widetilde{S}_{1}T_{0,1}=-T_{0,1}\widetilde{S}_{1},\ \ \widetilde{S}_{1}\widetilde{S}_{0}=-\widetilde{S}_{0}\widetilde{S}_{1}.
\]
Hence,
\[
\begin{aligned}
\overline{w_{0\pi,h}(0,0)}
&=\langle\widetilde{S}_{1}r_{0,h},\widetilde{S}_{1} f_{\pi,h}\rangle
=-w_{0\pi,h}(0,0),\\
\overline{w_{00,h}(0,1)}
&=\langle\widetilde{S}_{1}r_{0,h},\widetilde{S}_{1}(T_{0,1}f_{0,h})\rangle
=-w_{00,h}(0,1).
\end{aligned}
\]

The proposition then follows by taking
\[
s_h:=-i\,w_{00,h}(0,1)\in\mathbb R,
\qquad
\tau_h:=w_{0\pi,h}(0,0)\in i\mathbb R
\]
in \eqref{ematrx}.
\end{proof}

\subsection{Non-vanishing of the interaction coefficients}
In this subsection we complete the proof of the main theorem by computing the leading terms of  $s_{h},\tau_{h}$ and proving that they are nonzero for sufficiently small $h$. We first express these coefficients as surface integrals, up to exponentially smaller errors.
\begin{prop}\label{surf}
    Let $\epsilon>0$.
    \begin{equation}\label{defgamma}
    \Gamma_{\mathrm{h}}:=\{(x_{1}, x_{2})\in\mathbb{R}^{2}\mid x_{1}=\tfrac{\pi}{2},\ |x_{2}|<\epsilon\},  \ \ \   \Gamma_{\mathrm{v}}:=\{(x_{1}, x_{2})\in\mathbb{R}^{2}\mid x_{2}=\pi,\ |x_{1}|<\epsilon\}.
    \end{equation}
Then there exists $c\gg\eta>0$ such that
    \begin{equation}\label{surfaceintegrals}
   \begin{split}
       w_{0\pi, h}(0,0)&=h^{2}\int_{\Gamma_{\mathrm{h}}}\big(\langle\nabla^{k_{0}}_{\partial_{x_{1}}}f_{0, h}, f_{\pi, h}\rangle-\langle f_{0, h},\nabla^{k_{0}}_{\partial_{x_{1}}}f_{\pi, h}\rangle\big)\,dS_{g}+\mathcal{O}(e^{-(S+c)/h}),
       \\
       w_{00, h}(0, 1)&=h^{2}\int_{\Gamma_{\mathrm{v}}}\big(\langle\nabla^{k_{0}}_{2\partial_{x_{2}}}f_{0, h}, T_{0, 1}f_{0, h}\rangle-\langle f_{0, h}, \nabla^{k_{0}}_{2\partial_{x_{2}}}(T_{0, 1}f_{0, h})\rangle\big)\,dS_{g}+\mathcal{O}(e^{-(S+c)/h}).
   \end{split}     
    \end{equation}
\end{prop}
\begin{proof}
We prove the first formula in \eqref{surfaceintegrals}; the proof of the second is analogous. 
    
    Let
    \[
    \begin{split}
   D_{R}:=(\tfrac{\pi}{2}, R)\times(-R, R),\ \ \mathbb{H}_{+}:=\{(x_{1}, x_{2})\in\mathbb{R}^{2}\mid x_{1}>\tfrac{\pi}{2}\},
   \\
   \mathbb{H}_{-}:=\{(x_{1}, x_{2})\in\mathbb{R}^{2}\mid x_{1}<\tfrac{\pi}{2}\}.\hspace{17mm}
   \end{split}
    \]
For $u, v\in C^{\infty}(\overline{{D}_{R}};\mathbb{C}^{2}),$ we have
\[
\langle(\nabla^{k_{0}})^{*}\nabla^{k_{0}}u, v\rangle_{L^{2}(D_{R}; E)}=\langle \nabla^{k_{0}}u, \nabla^{k_{0}}v\rangle_{L^{2}(D_{R};T^{*}M\otimes E)}-\int_{\partial D_{R}}\langle\nabla^{k_{0}}_{\nu_{D}}u,v\rangle_{E_{x}}\,dS_{g},
\]
where $\nu_{D}\in C^{\infty}(\partial D_{R}; N\partial D_{R})$ is the outward unit normal. Hence
\[
\langle P_{h}(k_{0}) u, v\rangle_{L^{2}(D_{R})}-\langle u, P_{h}(k_{0})v\rangle_{L^{2}(D_{R})}=h^{2}\int_{\partial D_{R}}\big(\langle u, \nabla^{k_{0}}_{\nu_{D}}v\rangle-\langle\nabla^{k_{0}}_{\nu_{D}}u, v\rangle\big)\,dS_{g}.
\]
Taking $u=f_{0, h}$, $v=f_{\pi, h}$, and letting $R\rightarrow\infty,$ we can write
\begin{equation*}
\begin{split}
 w_{0\pi, h}(0,0)&:=\langle r_{0, h}, f_{\pi, h}\rangle_{L^{2}(\mathbb{R}^{2})} 
 \\
 &\ =\langle r_{0, h}, f_{\pi, h}\rangle_{L^{2}(\mathbb{H}_{-})}+\langle r_{0, h}, f_{\pi, h}\rangle_{L^{2}(\mathbb{H}_{+})}
 \\
 &\ =\langle r_{0, h}, f_{\pi, h}\rangle_{L^{2}(\mathbb{H}_{-})}+\langle f_{0, h}, r_{\pi, h}\rangle_{L^{2}(\mathbb{H}_{+})}
 \\
 &\ \ \ \ \ \ \ \ \ \ \ \ \ \ \ \ +h^{2}\int_{\{x_{1}=\frac{\pi}{2}\}}\big(\langle\nabla^{k_{0}}_{\partial_{x_{1}}}f_{0, h}, f_{\pi, h}\rangle-\langle f_{0, h},\nabla^{k_{0}}_{\partial_{x_{1}}}f_{\pi, h}\rangle\big)\,dS_{g}.
\end{split}
\end{equation*}
\eqref{surfaceintegrals} then follows from the Agmon estimates \eqref{pointwisedecay} and the fact that 
\[
d_{\widetilde{g}}(x_{0}, x)+d_{\widetilde{g}}(x_{\pi}, x)>S+c
\]
on $(\mathrm{supp}\, r_{0, h}\cap\mathbb{H}_{-})\cup (\mathrm{supp}\,r_{\pi, h}\cap \mathbb{H}_{+})$ for some $c\gg\eta>0.$
\end{proof}
We now derive the asymptotics of $w_{0\pi, h}(0,0), w_{00, h}(0, 1)$ by replacing $f_{0, h}, f_{\pi, h}$ by the corresponding normalized WKB approximate eigenfunctions in \eqref{surfaceintegrals}. More precisely, we have
\begin{prop}\label{comparisonwithwkb}
Let $\sigma\in\{0,\pi\},$ and let $U_{\sigma}\Subset \Omega_{\sigma}$ satisfy the assumptions \eqref{furtherassumption},\eqref{starshape}, with $\Phi_{\sigma}:=d_{\widetilde{g}}(x_{\sigma},\bullet)$ smooth. Let
\begin{equation}\label{normalizedwkbsolution}
 u_{\sigma, h}(x):=h^{-\frac{1}{2}}c_{\sigma}(h)e^{-\Phi_{\sigma}(x)/h}a_{\sigma}(x;h),\ \ a_{\sigma}(x;h)\sim\sum_{j=0}^{\infty}h^{j}a_{\sigma, j}(x),\ \ c_{\sigma}(h)=\mathcal{O}(1)
\end{equation}
be the corresponding WKB approximate eigenfunctions to \eqref{approximateeigenfunction} constructed in Proposition \ref{wkb}, so that $\|\chi_{\sigma}u_{\sigma, h}\|_{L^{2}}=1$ for a fixed $\chi_{\sigma}\in C^{\infty}_{c}(U_{\sigma};[0, 1])$ equal to 1 near $x_{\sigma}$. After multiplying normalized Dirichlet eigenfunctions $\phi_{\sigma, h}$ on $\Omega_{\sigma}$ by constants of modulus $1$, if necessary, one has, for every compact set $K\subset U_{\sigma},$ $\alpha\in\mathbb{N}_{0}^{2}$ and $N\in \mathbb{N}_{0},$
\begin{equation}\label{eq:agmonwkb}
\sup_{x\in K}e^{\Phi_{\sigma}(x)/h}|\partial^{\alpha}(u_{\sigma, h}-\phi_{\sigma, h})(x)|\leq C_{K, \alpha,N}h^{N}.    
\end{equation}
\end{prop}
%\begin{lemm}\label{sharpenedestimate}
%Suppose $v(h)\in H^{1}_{0}(U; E)\cap H^{2}(U; E)$ satisfies
%\[
%\begin{split}
%(P_{h, U}-\lambda(h))v(h)=0,\ \|v(h)\|_{L^{2}(U; E)}=1,
%\\
%\lambda(h)\in [0, Mh],\ M>0.\ \ \ \ \ \ \ 
%\end{split}
%\]
%There exist constant $N, C>0$, such that
%\begin{equation}
%h^{2}\|(1+\tfrac{\Phi}{h})^{-N}e^{\Phi/h}\nabla v\|^{2}_{L^{2}(U;T^{*}M\otimes E)}+h\|(1+\tfrac{\Phi}{h})^{-N}e^{\Phi/h}v\|^{2}_{L^{2}(U;E)}\leq Ch,    
%\end{equation}
%where $\Phi(x):=d_{\mathrm{Ag}}(x_{0}, x)$
%\end{lemm}
%\begin{proof}
%After replacing \cite[Proposition 6.2]{DiSj99} by Lemma \ref{agmonii}, the  proof proceeds line by line as in \cite[Proposition A.2]{DiSj99}.   
%\end{proof}
\begin{proof}
    We adapt the arguments of \cite[Chapter~6, Proposition~A.2 and Theorem~A.3]{DiSj99}. We remark that $U_{\sigma}$ can be chosen to contain the horizontal segment from $x_{\sigma}$ to $(\frac{\pi}{2}, 0)$ by Proposition \ref{geometricproperties}. 

   % 1. Proposition \ref{simiplicityofonewell}, as applied to $\Omega_{\sigma}$ in \eqref{dirichleteigenfunctiononcover} gives $\lambda_{\sigma,h}-h\Lambda(h)=\mathcal{O}(h^{\infty})$ and a spectral gap at least $c h$ about $\lambda_{\sigma, h}$, where $\lambda_{\sigma, h}$ are the Dirchlet eigenvalues corresponding to $U_{\sigma}.$ Hence, $\|(P_{h}(k_{0})-\lambda_{h})(\chi_{\sigma}u_{\sigma, h})\|_{L^{2}}=\mathcal{O}(h^{\infty}),$ a choice of phase of $\phi_{\sigma, h}$ yields
    %\begin{equation}\label{eq:wkbappro1}
   % \|\chi_{\sigma}u_{\sigma, h}-\phi_{\sigma, h}\|_{L^{2}}=\mathcal{O}(h^{\infty}).
   % \end{equation}
   1. We prove a sharpened Agmon estimate for $\phi_{\sigma, h}.$ The nondegeneracy of $\mu$ gives a constant $C_{0}>0$ such that ${\mu}/{\Phi_{\sigma}}\geq C_{0}.$ Choose $C>0$ such that $CC_{0}>2\Lambda_{0}+1$ and put
   \[
   \widetilde{\Phi}_{\sigma, h}:=\Phi_{\sigma}-Ch\log\max({\Phi_{\sigma}}/{h}, C).
   \]
   We then have
   \[
   \begin{split}
       \mu-|d\widetilde{\Phi}_{\sigma, h}|_{g}^{2}&=0\ \ \mathrm{a.e.\ on}\ \{x\in \Omega_{\sigma}\mid \Phi_{\sigma}(x)\leq Ch\},
       \\
       \mu-|d\widetilde{\Phi}_{\sigma, h}|^{2}_{g}&\geq CC_{0}h\ \ \mathrm{a.e.\ on}\ \{x\in\Omega_{\sigma}\mid\Phi_{\sigma}> Ch\}.
   \end{split}
   \]
   Let $a_{\sigma, h}:=\mu-|d\widetilde{\Phi}_{\sigma,h}|_{g}^{2}-\lambda_{h},$ where $\lambda_{h}$ denotes the Dirichlet eigenvalue on $\Omega_{\sigma}$ as in \eqref{dirichleteigenfunctiononcover}. We now take
   \[
    F_{\sigma,+}:=\sqrt{h}\,\textbf{1}_{\{\Phi_{\sigma}\leq Ch\}}+\sqrt{a_{\sigma, h}}\,\textbf{1}_{\{\Phi_{\sigma}> Ch\}},\ \ F_{\sigma, -}:=\sqrt{h+\lambda_{h}}\,\textbf{1}_{\{\Phi_{\sigma}\leq Ch\}}
   \]
and apply Lemma \ref{agmonii} to $P_{h}(k_{0})-\lambda_{h}$ and $\phi_{\sigma, h}$ to obtain 
\begin{equation*}
   \begin{split}
 h^2\|\nabla^{k_{0}} (e^{\widetilde{\Phi}_{\sigma, h}/h}\phi_{\sigma, h})\|_{L^{2}(\Omega_{\sigma})}^2+\tfrac12\|F_{\sigma,+}e^{\widetilde{\Phi}_{\sigma, h}/h}\phi_{\sigma, h}\|_{L^{2}(\Omega_{\sigma})}^2
 &\le\tfrac32\|F_{\sigma, -}e^{\widetilde{\Phi}_{\sigma, h}/h}\phi_{\sigma, h}\|_{L^{2}(\Omega_{\sigma})}^2\notag\\
 &=\tfrac32(h+\lambda_h)\|e^{\widetilde{\Phi}_{\sigma, h}/h}\phi_{\sigma, h}\|_{L^2(\{\Phi_{\sigma}\leq Ch\})}^2.
\end{split}
\end{equation*}
Since $F_{\sigma, +}\geq \sqrt{h}$, $\lambda_{h}\leq 2\Lambda_{0}h$,  there exists $\widetilde{C}>0$ such that
\[
h^{2}\|\nabla^{k_{0}}(e^{\widetilde{\Phi}_{\sigma, h}/h}\phi_{\sigma, h})\|^{2}_{L^{2}(\Omega_{\sigma})}+h\|e^{\widetilde{\Phi}_{\sigma, h}/h}\phi_{\sigma, h}\|_{L^{2}(\Omega_{\sigma})}^{2}\leq\widetilde{C}h.
\]
Hence, if $N_{0}\geq C+1,$
\begin{equation}\label{sharpenedwkb}
 \|e^{\Phi_{\sigma}/h}\phi_{\sigma, h}\|_{L^{2}(\Omega_{\sigma})}+\|e^{\Phi_{\sigma}/h}\nabla^{k_{0}}\phi_{\sigma, h}\|_{L^{2}(\Omega_{\sigma})}=\mathcal{O}(h^{-N_{0}}).   
\end{equation}

\ \ \ 2. For a fixed compact set $K\subseteq U_{\sigma},$ define
\[
\widehat{K}_{\sigma}:=\{x_{\sigma}\}\cup \{\exp(t\,\mathrm{grad}_{g}\Phi_{\sigma})(x)\mid x\in K,\ t\leq 0\}.
\]
Choose $\widetilde{\chi}_{\sigma}\in C^{\infty}_{c}(U_{\sigma};[0, 1])$, $\widetilde{\chi}_{\sigma}\equiv 1$ near $\widehat{K}_{\sigma},$ and set
\[
F_{\sigma}(x):=\min_{y\in\mathrm{supp}\,d\widetilde{\chi}_{\sigma}}(\Phi_{\sigma}(y)+d_{\widetilde{g}}(y, x)).
\]
Then $F_{\sigma}\geq\Phi_{\sigma}$ with equality on $\mathrm{supp}\,d\widetilde{\chi}_{\sigma}$, $|dF_{\sigma}|_{g}^{2}\leq\mu$ a.e., and $F_{\sigma}>\Phi_{\sigma}$ on $K.$ We now fix $\epsilon>0$ so small that $(1-\epsilon)F_{\sigma}>\Phi_{\sigma}$ near $K$, and enlarge the constant $C$ in Step 1 so that $C\geq\max\{e, \frac{2\Lambda_{0}+1}{\epsilon C_{0}}\}.$ We put
\[
\Psi_{\sigma, h}:=\min\{\widetilde{\Phi}_{\sigma, h}, (1-\epsilon)F_{\sigma}\}.
\]
On $\{\Phi_{\sigma}\leq Ch\}$, 
\[
\begin{split}
\widetilde{\Phi}_{\sigma, h}\leq Ch(1-\log C)\leq (1-\epsilon)F_{\sigma},
\end{split}
\]
On $\{\Phi_{\sigma}> Ch\},$
\[
\begin{split}
    \mu-|d\widetilde{\Phi}_{\sigma, h}|^{2}_{g}&\geq CC_{0}h,
    \\
    \mu-|d((1-\epsilon)F_{\sigma})|^{2}_{g}&\geq (1-(1-\epsilon)^{2})\mu\geq \epsilon CC_{0}h.
\end{split}
\]
Hence,
\begin{equation*}
 \mu-|d\Psi_{\sigma,h}|_g^2-\lambda_h
 \begin{cases}
 =-\lambda_h,&\text{a.e. on }\{\Phi_{\sigma}\leq Ch\},\\
 \ge h,&\text{a.e. on } \{\Phi_{\sigma}>Ch\}.
 \end{cases}
\end{equation*}
We now take
\[
\widetilde{F}_{\sigma, +, h}=\sqrt{h},\ \ \widetilde{F}_{\sigma,-, h}=\sqrt{h+\lambda_{h}}\textbf{1}_{\{\Phi_{\sigma}\leq Ch\}}.
\]
Let $w_{\sigma, h}:=u_{\sigma, h}-\phi_{\sigma, h}.$ By Lemma \ref{agmonii},
\begin{equation*}
\begin{aligned}
&h^2\left\|
\nabla^{k_0}\!\left(e^{\Psi_{\sigma,h}/h}\widetilde{\chi}_\sigma w_{\sigma,h}\right)
\right\|_{L^2(\Omega_\sigma)}^2
+\tfrac{h}{2}\left\|
e^{\Psi_{\sigma,h}/h}\widetilde{\chi}_\sigma w_{\sigma,h}
\right\|_{L^2(\Omega_\sigma)}^2
\\
&\quad\le
\underbrace{
h^{-1}\left\|
e^{\Psi_{\sigma,h}/h}
(P_h(k_0)-\lambda_h)(\widetilde{\chi}_{\sigma} w_{\sigma,h})
\right\|_{L^2(\Omega_\sigma)}^2
}_{:=R_{1,\sigma,h}}
+
\underbrace{
\tfrac{3}{2}(h+\lambda_h)\left\|
e^{\Psi_{\sigma,h}/h}\widetilde{\chi}_\sigma w_{\sigma,h}
\right\|_{L^2(\{\Phi_\sigma\le Ch\})}^2
}_{:=R_{2,\sigma,h}}.
\end{aligned}
\end{equation*}
After modifying $\phi_{h,\sigma}$ by a constant of modulus 1, we have
\[
R_{2,\sigma, h}\lesssim h\|\chi_{\sigma}u_{\sigma, h}-\phi_{\sigma, h}\|^{2}_{L^{2}(\Omega_{\sigma})}=\mathcal{O}(h^{\infty}).
\]
To estimate $R_{1,\sigma, h},$ we write 
\[
(P_{h}(k_{0})-\lambda_{h})(\widetilde{\chi}_{\sigma}w_{\sigma, h})=\widetilde{\chi}_{\sigma}(P_{h}(k_{0})-\lambda_{h})w_{\sigma, h}+[P_{h}(k_{0}),\widetilde{\chi}_{\sigma}]w_{\sigma, h}.
\]
The first term is $\mathcal{O}(h^{\infty})$ in the weighted norm. Since $\Psi_{\sigma, h}=(1-\epsilon)\Phi_{\sigma}$ on $\mathrm{supp}\,d\widetilde{\chi}_{\sigma},$ we have by \eqref{sharpenedwkb}
\[
\|e^{\Psi_{\sigma, h}/h}[P_{h}(k_{0}), \widetilde{\chi}_{\sigma}]w_{\sigma, h}\|_{L^{2}(\Omega_{\sigma})}\lesssim h^{-M}e^{-\epsilon\min_{\mathrm{supp}\,d\widetilde{\chi}_{\sigma}}\Phi_{\sigma}/h}=\mathcal{O}(h^{\infty}).
\]
Thus, $\|e^{\Phi_{\sigma}/h}w_{\sigma, h}\|_{L^{2}(K)}=\mathcal{O}(h^{\infty})$ and \eqref{eq:agmonwkb} follows from elliptic regularity.

\end{proof}
\begin{prop}\label{Prop:nonvanishing}
    There exists $\mathfrak{c}>0$ such that
    \begin{equation}\label{consts}
    \begin{split}
    |w_{0\pi, h}(0,0)|&=\mathfrak{c}h^{\frac{1}{2}}e^{-S/h}+\mathcal{O}(h^{\frac{3}{2}}e^{-S/h}),
    \\
    |w_{00, h}(0, 1)|&=\mathfrak{c}h^{\frac{1}{2}}e^{-S/h}+\mathcal{O}(h^{\frac{3}{2}}e^{-S/h}).
    \end{split}
    \end{equation}
\end{prop}
\begin{proof}
   By Propositions \ref{surf} and \ref{comparisonwithwkb},
    \[
    \begin{split}
    |w_{0\pi, h}(0,0)|&=\tfrac{1}{2}h^{2}\bigg|\int_{\Gamma_{\mathrm{h}}}\big(\langle\nabla^{k_{0}}_{\partial_{x_{1}}}u_{0, h}, u_{\pi, h}\rangle-\langle u_{0, h}, \nabla^{k_{0}}_{\partial_{x_{1}}}u_{\pi, h}\rangle\big)\,\mathrm{d}x_{2}\bigg|+\mathcal{O}(h^{\infty}e^{-S/h})
    \\
    &=\tfrac{1}{2}|c_{0}(h)\overline{c_{\pi}(h)}|\bigg|\int_{\Gamma_{\mathrm{h}}}e^{-(\Phi_{0}+\Phi_{\pi})/h}(\partial_{x_{1}}\Phi_{\pi}-\partial_{x_{1}}\Phi_{0})\langle a_{0}, a_{\pi}\rangle\,\mathrm{d}x_{2}
\\
    &\hspace{15mm}+h\int_{\Gamma_{\mathrm{h}}}e^{-(\Phi_{0}+\Phi_{\pi})/h}\big(\langle\nabla^{k_{0}}_{\partial_{x_{1}}}a_{0}, a_{\pi}\rangle-\langle a_{0}, \nabla^{k_{0}}_{\partial_{x_{1}}}a_{\pi}\rangle\big)\,\mathrm{d}x_{2}\bigg|+\mathcal{O}(h^{\infty}e^{-S/h}).
    \end{split}
    \]
    Let $|c_{0}(h)\overline{c_{\pi}(h)}|=c+\mathcal{O}(h)$. We apply Proposition \ref{geometricproperties} and \cite[Theorem 7.7.5]{Hormander I} to obtain
    \[
    |w_{0\pi, h}(0,0)|=\tfrac{c}{2}e^{-S/h}\big(\frac{2\pi h}{\Psi''(0)}\big)^{\frac{1}{2}}\Big(\Big|(\partial_{x_{1}}\Phi_{\pi}-\partial_{x_{1}}\Phi_{0})\langle a_{0,0}, a_{\pi, 0}\rangle(\tfrac{\pi}{2}, 0)\Big|+\mathcal{O}(h)\Big)+\mathcal{O}(h^{\infty}e^{-S/h}).
    \]

Since $\partial_{x_{1}}\Phi_{\pi}(\frac{\pi}{2}, 0)=-\partial_{x_{1}}\Phi_{0}(\frac{\pi}{2},0)$ and $a_{0,0}(\frac{\pi}{2},0), a_{\pi, 0}(\frac{\pi}{2}, 0)$ are solutions of transport equations \eqref{scalarhomogeneoustransport} in the same one-dimensional eigenspace, it follows that
\[
(\partial_{x_{1}}\Phi_{\pi}-\partial_{x_{1}}\Phi_{0})\langle a_{0,0}, a_{\pi, 0}\rangle(\tfrac{\pi}{2}, 0)\neq 0.
\]
It remains to justify the agreement of the leading coefficients. This follows from the fact that if we write
\[
a_{\sigma, 0}=b_{\sigma}e_{\sigma},\ X_{\sigma}=\mathrm{grad}_{g}\Phi_{\sigma}
\]
with $|e_{\sigma}|=1$, then the connection form is purely imaginary, so multiplying the scalar transport equation by \(\overline{b_\sigma}\) and taking real parts gives
\begin{equation}\label{eq:transportequationb}
X_{\sigma}|b_{\sigma}|^{2}+(\Delta_{g}\Phi_{\sigma}-\Lambda_{0})|b_{\sigma}|^{2}=0.
\end{equation}
The isometry $(x_{1}, x_{2})\mapsto(x_{2}/2, 2x_{1})$ preserves the equation and interchanges the horizontal and vertical pairs of wells. The proof is complete.
\end{proof}
We now combine these tunneling asymptotics with the symmetries of the model to show that the two lowest bands meet conically at
$k_0=(\tfrac12,\tfrac12)$.
\begin{prop}\label{prop:dirac}
    Let $E_{D}(h):=E_{1}(k_{0})=E_{2}(k_{0}).$ For a fixed and sufficiently small $h>0$,
   \begin{equation}\label{dirac:cone}
E_{1,2}(k_0+\kappa)
=E_D(h)\mp\alpha(h)|\kappa|+\mathcal O_h(|\kappa|^2),
\qquad \kappa\longrightarrow0,
\end{equation}
where $\alpha(h)>0$. 
More precisely, if $\mathfrak c>0$ is the constant in Proposition~\ref{Prop:nonvanishing}, then
\begin{equation}\label{dirac:slope-asymptotic}
\alpha(h)=2\pi\mathfrak c\,h^{1/2}e^{-S/h}\bigl(1+\mathcal O(h)\bigr).
\end{equation}
\end{prop}
\begin{proof}
    1. Proposition \ref{globaloperatorspectralgap} and \cite[Proposition~4.1, (4.6)]{GuZw26} give the formula
    \begin{equation}\label{dirac:gz}
\begin{gathered}
E_{1,2}(k_0+\kappa)
=E_D(h)\mp\sqrt{q_h(\kappa)}+\mathcal O_h(|\kappa|^2),\\
q_h(\kappa)=A(h)\kappa_1^2+2C(h)\kappa_1\kappa_2+D(h)\kappa_2^2\geq0.
\end{gathered}
\end{equation}
We fix $B=\pi$ and define unitary operators
\[
\begin{split}
(\mathcal{S}_{1}(k_{2})u)(x_{1}, x_{2})&:=e^{i(x_{1}-\pi k_{2})}u(2\pi k_{2}-x_{1}, x_{2}),
\\
(\mathcal{S}_{2}(k_{1}, k_{2})u)(x_{1}, x_{2})&:=\tfrac{1}{2}[u(x_{2}+\pi k_{2}, x_{1}-\pi k_{1})+u(x_{2}+\pi k_{2}+\pi, x_{1}-\pi k_{1})
\\
&\ \ \ \ \ \ \ \ \ \ \  \ +u(x_{2}+\pi k_{2}, x_{1}-\pi k_{1}+\pi)-u(x_{2}+\pi k_{2}+\pi, x_{1}-\pi k_{1}+\pi)].
\end{split}
\]
A computation shows that
\[
\begin{split}
\mathcal{S}_{1}(k_{2})H(k_{1}, k_{2})\mathcal{S}_{1}(k_{2})^{*}&=H(1-k_{1}, k_{2}),
\\
\mathcal{S}_{2}(k_{1}, k_{2})H(k_{1}, k_{2})\mathcal{S}_{2}(k_{1}, k_{2})^{*}&=H(k_{2}, k_{1}).
\end{split}
\]
Hence, $q_h ( \kappa_1, \kappa_2  ) = q_h ( \kappa_2, \kappa_1) =  q_h ( - \kappa_1, \kappa_2),$
and consequently, $ A(  h ) = D (h)$, $ C (h) = 0$.
This establishes \eqref{dirac:cone} with $\alpha(h)\geq 0.$

2. To compute $\alpha(h)$ using WKB approximations,  we now show the identity
\begin{equation}\label{eq:alpha}
   \alpha(h)=2h^{2}|\langle (D_{x_{1}}-\tfrac{1}{2})e_{0}, e_{\pi}\rangle|,
\end{equation}
where $e_{\sigma}:=e_{\sigma, h}(0), \sigma\in\{0,\pi\}$ is the orthonormal basis defined by \eqref{orthonormalize}.

Let $\mathcal{U}_{0}$ be the unitary reduction of Proposition \ref{reductionatoddpi} at $k_{0}$, and let $\psi_{\pm}$ be defined by \cite[Proposition~4.1, (4.5)]{GuZw26}. We record the identities
\[
\mathcal{U}_{0}(D_{x_{1}}-\tfrac{1}{2})\mathcal{U}_{0}^{*}=(D_{x_{1}}-\tfrac{1}{2})I_{\mathbb{C}^{2}},\ \ \ \mathcal{U}_{0}\psi_{\pm}=e^{i\theta_{\pm}}u_{\pm},
\]
where $u_{\pm}:=\frac{1}{\sqrt{2}}(e_{0}\pm e_{\pi}).$ Hence, \cite[Proposition~4.1, (4.7)]{GuZw26} gives
\[
\alpha(h)=2h^{2}|\langle (D_{x_{1}}-\tfrac{1}{2})u_{+}, u_{+}\rangle|.
\]
After conjugation, the identities
\[
[D_{x_{1}}-\tfrac{1}{2}, S_{0}]=0,\ \ (D_{x_{1}}-\tfrac{1}{2})S_{1}=-S_{1}(D_{x_{1}}-\tfrac{1}{2})
\]
used in \cite[Proposition~4.1]{GuZw26} give
\[
\langle (D_{x_{1}}-\tfrac{1}{2})u_{+}, u_{-}\rangle=0,\ \ \langle(D_{x_{1}}-\tfrac{1}{2})u_{-}, u_{-}\rangle=-\langle(D_{x_{1}}-\tfrac{1}{2})u_{+}, u_{+}\rangle.
\]

Therefore,
\[
\begin{aligned}
\left\langle
\left(D_{x_1}-\tfrac{1}{2}\right)e_0,e_\pi
\right\rangle
&=\frac{1}{2}\left\langle
\left(D_{x_1}-\tfrac{1}{2}\right)(u_++u_-),u_+-u_-
\right\rangle\\
&=\left\langle
\left(D_{x_1}-\tfrac{1}{2}\right)u_+,u_+
\right\rangle.
\end{aligned}
\]

3. Using the metric and connection in \eqref{metric,connection}, set $\Sigma_{s}:=\{(s, y)\mid -\pi\leq y\leq \pi\}$, and define
\[
J_{h}(s):=h^{2}\int_{\Sigma_{s}}\big(\langle\nabla^{k_{0}}_{\partial_{x_{1}}}e_{0}, e_{\pi}\rangle-\langle e_{0},\nabla^{k_{0}}_{\partial_{x_{1}}}e_{\pi}\rangle\big)\,dS_{g}.
\]

Since $P_{h}(k_{0})=-h^{2}(\nabla_{\partial_{x_{1}}}^{k_{0}})^{2}-4h^{2}(\nabla^{k_{0}}_{\partial_{x_{2}}})^{2}+V,$
\[
\begin{split}
J'_{h}(s)&=h^{2}\int_{\Sigma_{s}}\big(\langle(\nabla^{k_{0}}_{\partial_{x_{1}}})^{2}e_{0}, e_{\pi}\rangle-\langle e_{0},(\nabla^{k_{0}}_{\partial_{x_{1}}})^{2}e_{\pi}\rangle\big)\,dS_{g}
\\
&=\int_{\Sigma_{s}}\langle e_{0}, P_{h}(k_{0})e_{\pi}\rangle-\langle P_{h}(k_{0})e_{0}, e_{\pi}\rangle\,dS_{g}
=0.
\end{split}
\]
Hence,
\[
\alpha(h)=\Big|\int_{-\pi}^{\pi} J_{h}(s)\,\mathrm{d}s\Big|=2\pi |J_{h}(\tfrac{\pi}{2})|.
\]
The proof of Proposition \ref{asymptoticeffecitivematrix} gives
\[
J_{h}(\tfrac{\pi}{2})=h^{2}\int_{\Sigma_{s}}\big(\langle\nabla^{k_{0}}_{\partial_{x_{1}}}f_{0, h}, f_{\pi, h}\rangle-\langle f_{0, h},\nabla^{k_{0}}_{\partial_{x_{1}}}f_{\pi, h}\rangle\big)\,dS_{g}+\widetilde{\mathcal{O}}(e^{-3S/2h}).
\]
Consequently, we can apply Propositions \ref{surf} and \ref{Prop:nonvanishing} to conclude \eqref{dirac:slope-asymptotic}. The proof is complete.
\end{proof}

%%%%%%%%%%%%%%%%%%%%%%%%%%%%%%%%%%%%%%%%%%%%%%%%%%%%%%%%%%%%%%%%%%%%%%
\section*{Appendix. Numerical Results}\label{Appendix}
\begingroup
\renewcommand{\thesubsection}{A.\arabic{subsection}}
\renewcommand{\theHsubsection}{appendix.\arabic{subsection}}
\renewcommand{\theequation}{A.\arabic{equation}}
\renewcommand{\theHequation}{appendix.\arabic{equation}}
\setcounter{subsection}{0}
\setcounter{equation}{0}
\subsection{Numerical evaluation of \(C\)}
We compute the constant $C$ in \eqref{globalasymptotics}
and compare the leading formula with numerical band splitting at $B=\pi$.
We then examine the gap for $0\leq B<\pi$, providing numerical evidence
for Conjecture~\ref{app:conjecture}.

Set
$x=x_1$, $y=x_2/2$, and put
\[
 L:=\frac\pi2,\qquad \nu:=2^{-3/4},\qquad
 \rho(x):=\sqrt{1+\cos^2x},\qquad
 p(x):=\frac{\sin x}{\sqrt{\sqrt2+\rho(x)}}.
\]
For $\phi_{0}(x,y):=\Phi_0(x,2y)$, we have $\phi_{0,x}(x,0)=p(x)$.
By evenness in $y$, differentiating the eikonal equation
$\phi_{0,x}^2+\phi_{0,y}^2=\sqrt2-\sqrt{\cos^2x+\cos^2y}$ twice in $y$
at $y=0$ gives the equation for $r(x):=\phi_{0,yy}(x,0)$:
\begin{equation}\label{app:riccati}
 pr'+r^2=\frac1{2\rho},\qquad r(0)=\nu.
\end{equation}
The initial value follows from
$\phi_{0}(x,y)=\frac\nu2(x^2+y^2)+\mathcal O(|(x,y)|^4)$.

We write \eqref{normalizedwkbsolution} in the $(x,y)$ coordinates as
\[
 u_{\sigma,h}=h^{-1/2}e^{-\phi_\sigma/h}
 (\alpha_\sigma+\mathcal O(h)).
\]
Let $A(x):=|\alpha_0(x,0)|.$ Then
\[
\begin{split}
    1=\|\chi u_{0, h}\|^{2}_{L^{2}}&=h^{-1}\int_{\mathbb{R}^{2}}\chi^{2}e^{-2\phi_{0}/h}\big(|\alpha_{0}|^{2}+\mathcal{O}(h)\big)\,\mathrm{d}x\mathrm{d}y
    \\
    &=A(0)^{2}\int_{\mathbb{R}^{2}}e^{-\nu(x^{2}+y^{2})}\,\mathrm{d}x\mathrm{d}y.
\end{split}
\]
Consequently, $A(0)^{2}={\nu}/{\pi}.$ On $y=0$, we have $X_{0}=p\partial_{x}$, $\Delta_{g}\phi=p'+r$ and $\Lambda_{0}=2\nu.$ Thus, \eqref{eq:transportequationb} gives
\[
p(A^{2})'+(p'+r-2\nu)A^{2}=0,\ \ \frac{d}{dx}\log\big(\frac{pA^{2}}{x}\big)=\frac{2\nu-r}{p}-\frac{1}{x}.
\]
Since $p(x)=\nu x+\mathcal{O}(x^{3}),$ we obtain
\[
p(L)A(L)^{2}=\frac{\nu^{2}L}{\pi}\exp\bigg(\int_{0}^{L}\Big(\frac{2\nu-r(x)}{p(x)}-\frac{1}{x}\Big)\,\mathrm{d}x\bigg).
\]
The proof of Proposition \ref{Prop:nonvanishing} and \eqref{globalasymptotics1} give
\[
C=4\mathfrak{c}=8p(L)A(L)^{2}\sqrt{\frac{\pi}{r(L)}}.
\]
For the numerical evaluation, we use the expansions
\[
r(x)=\nu\left(1+\frac{x^2}{16}+\mathcal O(x^4)\right),
\qquad
\frac{2\nu-r(x)}{p(x)}-\frac1x
=\frac{x}{24}+\mathcal O(x^3),
\]
which follow from (A.1). In particular, the integrand extends smoothly to \(x=0\). Numerically solving (A.1) and evaluating the integral gives
\[
r(L)\simeq0.677623467661,
\qquad
\int_0^L\left(\frac{2\nu-r(x)}{p(x)}-\frac1x\right)\,dx
\simeq0.0797616804392.
\]
Since \(p(L)=\sqrt{\sqrt2-1}\), the preceding identities yield
\[
A(L)^2\simeq0.297476669937,
\qquad
C\simeq3.29788701535.
\]

We also record the value of the tunneling constant given in \eqref{tunnelingdistance}.
\[
S=2\int_0^L p(x)\,dx\simeq1.250624857493.
\]
\par Using these values of $C$ and $S$, Figure~\ref{app:fig-pi} compares the
numerical band splitting with the leading term of \eqref{globalasymptotics}
along $k_1=k_2=t$:
\begin{equation}\label{app:scaled-gap}
\begin{gathered}
 a(h):=\sqrt h\,e^{-S/h},\\
 D_h(t,t)=\frac{E_2(t,t)-E_1(t,t)}{a(h)}
 \quad\text{compared with}\quad C\sqrt2|\cos(\pi t)|.
\end{gathered}
\end{equation}
The numerical curve in Figure~\ref{app:fig-pi} uses Fourier modes
$|m|,|n|\leq32$ at $t=j/32$, $j=0,\ldots,32$. The curves illustrate the size and quasi-momentum dependence predicted by
the main theorem.

\begin{figure}[p]
\centering
\includegraphics[width=\textwidth]{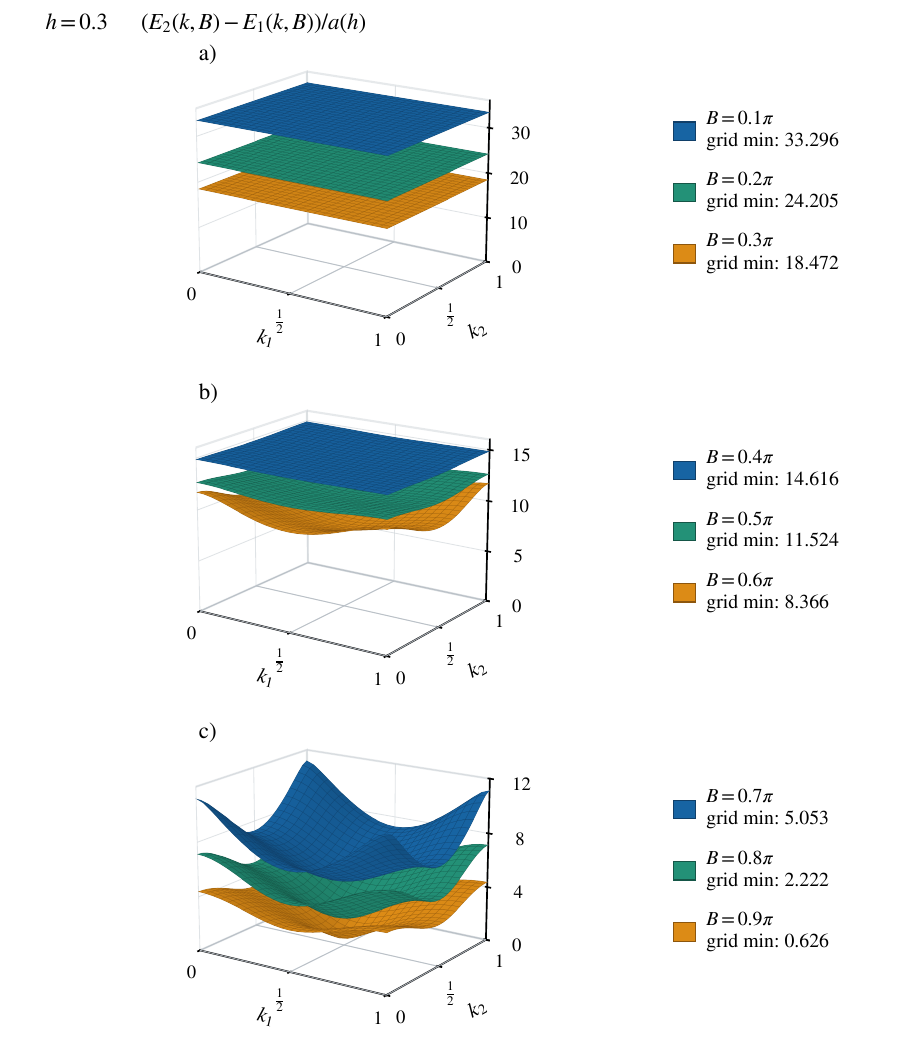}
\caption{The rescaled gap $D_h(k,B)$ on $[0,1]^2$ at $h=0.3$.
The three panels group $B/\pi=0.1,0.2,0.3$;
$0.4,0.5,0.6$; and $0.7,0.8,0.9$, respectively.
Each panel has its own vertical scale and includes the zero plane.
Legend entries give the sampled minima. Click \href{https://general-b-band-geometry.fiberinfinity.chatgpt.site/}
{here} for an interactive three-dimensional plot.}
\label{app:fig-surfaces}
\end{figure}

\subsection{Band splitting for general values of \(B\)}
\label{app:general-B}
Having compared the leading formula at $B=\pi$, we now examine the
separation of the first two bands for other values of $B$.
We use the same expression \eqref{originaloperator} for arbitrary
$B\in\mathbb R$, denote this family by $H_B(k)$, and write its
eigenvalues as $E_j(k;B,h)$. We set
\[
 D_h(k,B):=\frac{E_2(k;B,h)-E_1(k;B,h)}{a(h)},
\]
where $ a  ( h) $ is given in \eqref{app:scaled-gap}.

\par The middle identity in \eqref{eq:conjugationgeneralB} shows that the
conjugated operator depends on $B$ through $e^{\pm iBn}$,
$n\in\mathbb Z$, giving $E_j(k;B+2\pi,h)=E_j(k;B,h)$. The antiunitary involution \(Ju(x)=\overline{u(-x)}\) satisfies \(JH_B(k)J=H_{-B}(k)\), so \(E_j(k;-B,h)=E_j(k;B,h)\). Together with \(2\pi\)-periodicity, this shows that it suffices to consider \(0\leq B\leq\pi\).

We now compute the rescaled gap and examine its
closing as $B$ approaches $\pi$. For Figures~\ref{app:fig-surfaces}
and~\ref{app:fig-closing}, we restrict the Fourier matrix of the conjugated
operator in \eqref{eq:conjugationgeneralB} to $|m|,|n|\leq N$ and compute
its lowest two eigenvalues. Figure~\ref{app:fig-surfaces} uses $h=0.3$,
$N=16$, and a $49\times49$ grid in $[0,1]^2$ for $B/\pi=0.1,0.2,\ldots,0.9$.

\begin{figure}[!ht]
\centering
\includegraphics[width=11.5cm]{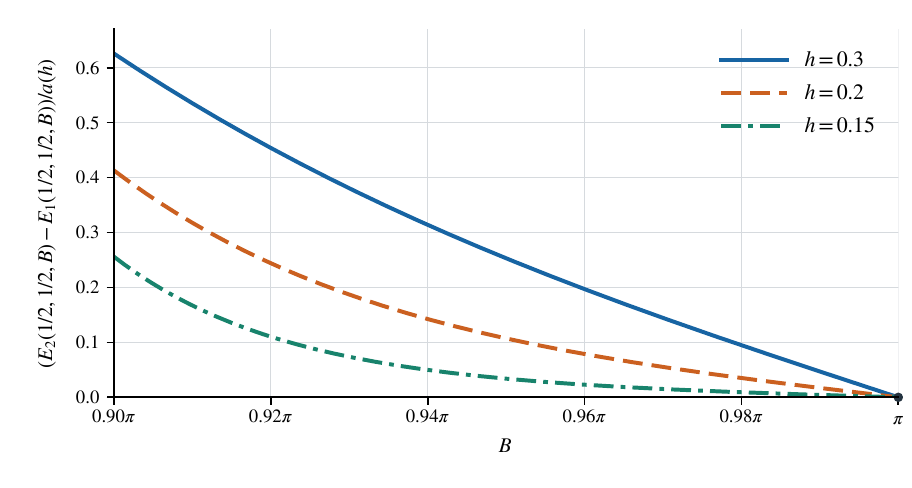}
\caption{The rescaled gap $D_h((1/2,1/2),B)$ for
$0.9\pi\leq B\leq\pi$ at $h=0.3,0.2,0.15$.
Each curve is divided by its own $a(h)=\sqrt h\,e^{-S/h}$
and has the exact symmetry-enforced zero at $B=\pi$.}
\label{app:fig-closing}
\end{figure}
%\FloatBarrier

We remark that at $B=0$, separation of variables gives
\[
 E_2(k;0,h)-E_1(k;0,h)
 =\min_{j=1,2}\bigl(\varepsilon_2(k_j)-\varepsilon_1(k_j)\bigr),
\]
where $\varepsilon_\ell(\kappa)$ are the eigenvalues of
$q_h(\kappa)=h^2(D_x-\kappa)^2+\cos x$. We can then apply \cite[\S6.3, Theorem~7 and Corollary~6.6]{TaZw23} to show that $$E_{2}(k;0,h)-E_1(k;0,h)=\sqrt{2}h+\mathcal{O}(h^{2}),$$ uniformly in $k$, while the band widths are exponentially small. Hence, near $B=0$ the rescaled bands appear nearly flat.

To examine the closing in more detail, Figure~\ref{app:fig-closing} fixes
$k=(1/2,1/2)$ and samples 301 equally spaced values of $B$ in
$[0.9\pi,\pi]$. We use $h=0.3,0.2,0.15$ with cutoffs $N=16,18,20$,
respectively, and divide each gap by the corresponding $a(h)$. The figure shows a positive gap at every sampled
$B<\pi$, closing at $\pi$. Together with
Figure~\ref{app:fig-surfaces}, this provides numerical evidence for
the following conjecture.

\begin{conj}\label{app:conjecture}
For every $ h > 0 $ and every $B\in[0,\pi)$,
\[
 E_2(k;B,h)>E_1(k;B,h)
 \qquad\text{for all }k\in\mathbb R^2/\mathbb Z^2.
\]
\end{conj}
This holds for sufficiently large $ h $ due to the perturbative result in \cite[Theorem 2]{GuZw26}.

\subsection{Clustering at rational values of \texorpdfstring{$B/(2\pi)$}{B/(2 pi)}}
\label{app:rational-B}
Fix $B=2\pi p/q$, with $(p,q)=1$ and $q\geq1$, and set $L_q=2\pi/q$.
We explain the clustering of the first $q$ bands in Figure~\ref{app:fig-rational}.

\emph{1. Reduction.}
Let $\mathsf S:\mathbb C^q\to\mathbb C^q$ be the cyclic shift
$(\mathsf Sv)_r=v_{(r+1)\bmod q}$.
The unitary map
\[
(\mathcal U_k u)_r(x,t)=u(x+Bk_2,t+rL_q),
\qquad 0\leq r<q,
\]
identifies $H_B(k)$ with the differential operator
\begin{equation*}\label{app:rational-operator}
\begin{aligned}
\mathcal P_h(k)
&=-h^2\bigl((\partial_x-ik_1)^2+(\partial_t-ik_2)^2\bigr)I_q
+\mathcal V(x,t),\\
\mathcal V(x,t)
&=\operatorname{diag}_{0\leq r<q}\cos(t+rL_q)
+\frac12\bigl(e^{ix}\mathsf S^p+e^{-ix}\mathsf S^{-p}\bigr).
\end{aligned}
\end{equation*}
It acts on sections of the rank-$q$ vector bundle
$\mathcal E_q\to M_q$ defined by
\[
M_q=(\mathbb R/2\pi\mathbb Z)\times(\mathbb R/L_q\mathbb Z),
\qquad
\mathcal E_q=(\mathbb R^2\times\mathbb C^q)/{\sim},
\]
where
\[
(x,t,z)\sim(x+2\pi m,t+nL_q,\mathsf S^n z),
\qquad m,n\in\mathbb Z.
\]
%The sections are represented by functions $v:\mathbb R^2\to\mathbb C^q$
%satisfying
%\[
%v(x+2\pi,t)=v(x,t),
%\qquad
%v(x,t+L_q)=\mathsf Sv(x,t).
%\]
\emph{2. Verification of \eqref{furtherassumption}.}
Let $e_{0}$ be the lowest eigenvalue of 
\[
\mathcal V(\pi,\pi)=-\operatorname{diag}_{0\leq r<q}\cos(rL_q)
-\tfrac12(\mathsf S^p+\mathsf S^{-p}).
\]
The simplicity of \(e_0\) is a consequence of the Perron–Frobenius theorem applied to the irreducible and nonnegative matrix \(2I_q-\mathcal V(\pi,\pi)\).

Define
\[
d(E;x,t):=\det(EI_q-\mathcal V(x,t)).
\]
Expanding the determinant, we obtain (see \cite[(3.4)]{HKR16})
\begin{equation}\label{app:rational-determinant}
d(E;x,t)
=F_{p,q}(E)-2^{1-q}\bigl(\cos(qx)+\cos(qt)\bigr),
\end{equation}
where $F_{p,q}$ is a monic polynomial of degree $q$, independent of $(x, t).$ If we denote the eigenvalues of $\mathcal{V}(\pi, \pi)$ by $\lambda_{1},\dots,\lambda_{q}$, where $\lambda_{1}=e_{0},$ then
\begin{equation}\label{eq:detidentity}
(-1)^{q}d(E;x, t)=\prod_{l=1}^{q}(\lambda_{l}-E)+2^{1-q}\big[2-(-1)^{q}\big(\cos(qx)+\cos(qt)\big)\big].
\end{equation}
For every $E<e_{0},$ the product term is strictly positive and the bracketed term is nonnegative. Thus,
\[
\mu(x, t):=\min\Spec\mathcal{V}(x, t)-e_{0}\geq 0.
\]
Evaluating the identity \eqref{eq:detidentity} at $E=e_0$ gives precisely $q$ wells:
\begin{equation*}\label{app:rational-wells}
\mu^{-1}(0)=\{z_j=[(\pi+jB,\pi)]\in M_q:0\leq j<q\}.
\end{equation*}
Differentiating
\eqref{app:rational-determinant} gives
\begin{equation*}\label{app:rational-hessian}
d\mu(z_j)=0,\qquad
\operatorname{Hess}_{z_j}\mu=
\frac{2^{1-q}q^2}{|F'_{p,q}(e_{0})|}I_2\gg0.
\end{equation*}
Hence the shifted potential $\mathcal V-e_0I_q$ satisfies
\eqref{furtherassumption} near each well.

\emph{3. Equivalent wells and clustering.}
For $\mathsf D=\operatorname{diag}_{0\leq r<q}e^{-irL_q}$,
\begin{equation*}\label{app:rational-symmetry}
\mathsf D\mathsf S^p\mathsf D^*=e^{iB}\mathsf S^p,
\qquad
\mathcal V(x+B,t)=\mathsf D\mathcal V(x,t)\mathsf D^*.
\end{equation*}
Thus $u(x,t)\mapsto\mathsf Du(x-B,t)$ identifies the Dirichlet
operators on translated small balls about the $z_j$.
The comparison argument of Proposition~\ref{globaloperatorspectralgap}, applied to these $q$ wells,
gives, for some $c_{p/q}>0$,
$E_j(k;B,h)=e(h)+\mathcal O(e^{-c_{p/q}/h})$ for $1\leq j\leq q$.
The WKB constructions give $e(h)=e_0+h\omega_{p/q}+\mathcal O(h^2)$ and $E_{q+1}(k;B,h)=e_0+2h\omega_{p/q}+o(h)$, where $\omega_{p/q}:=q\sqrt{\frac{2^{2-q}}{|F'_{p,q}(e_{0})|}}.$
Consequently, uniformly in $k$ as $h\to0$,
\begin{equation*}\label{app:rational-clustering}
\begin{aligned}
E_q(k;B,h)-E_1(k;B,h)&=\mathcal O(e^{-c_{p/q}/h}),\\
E_{q+1}(k;B,h)-E_1(k;B,h)&=h\omega_{p/q}+o(h).
\end{aligned}
\end{equation*}
%%%%%%%%%%%%%%%%%%%%%%%%%%%%%%%%%%%%%%%%%%%%%%%%%%%%%%%%%%%%%%%%%%%%

\endgroup

\end{document}